\documentclass[12pt,a4paper]{article}
\usepackage{jheppub_modified}

\allowdisplaybreaks[1]
\usepackage{amsmath,bm}
\usepackage{subfigure}

\title{Floquet superconductor in holography}

\author[1]{Takaaki Ishii}
\author[2]{and Keiju Murata}
\affiliation[1]{Institute for Theoretical Physics, Utrecht University, 3584 CC Utrecht, The Netherlands}
\affiliation[2]{Department of Physics, Osaka University, Toyonaka, Osaka 560-0043, Japan}
\emailAdd{t.ishii@uu.nl}
\emailAdd{murata@het.phys.sci.osaka-u.ac.jp}

\abstract{%
We study nonequilibrium steady states in a holographic superconductor under time periodic driving by an external rotating electric field.
We obtain the dynamical phase diagram.
Superconducting phase transition is of first or second order depending on the amplitude and frequency of the external source.
The rotating electric field decreases the superconducting transition temperature.
The system can also exhibit a first order transition inside the superconducting phase.
It is suggested this transition exists all the way down to zero temperature.
The existence of nonequilibrium thermodynamic potential for such steady solutions is also discussed from the holographic point of view.
The current induced by the electric field is decomposed into normal and superconducting components,
and this makes it clear that the superconducting one dominates in low temperatures.
}

\preprint{OU-HET-968}

\begin{document}
\maketitle

\section{Introduction}
\label{intro}

Understanding nonequilibrium dynamics in strongly correlated quantum many-body systems is one of the most significant problems 
in condensed matter physics.
A central goal in the study of non-equilibrium processes is to control the phases of matters by time-dependent external fields ~\cite{Review1,Review2}.
An example is to apply a laser with time periodicity to a material.  
If the energy injection balances the dissipation, a nonequilibrium steady state can be realized. 
Such a state in the presence of a periodic driving is called a {\it Floquet state}.

In this paper, we study Floquet states in a superconductor using the tools 
of the AdS/CFT correspondence, also known as holography~\cite{Maldacena:1997re,Gubser:1998bc,Witten:1998qj}.
Superconductivity is modeled in holography in Refs.~\cite{Gubser:2008px,Hartnoll:2008vx,Hartnoll:2008kx}.
We drive the holographic superconductor by a rotating electric field and investigate the phase structure of the Floquet states.

In the study of Floquet superconductors, 
one of the most striking consequences is the ``enhancement'' of superconductivity:
Irradiation of a laser to superconductors can increase the transition temperature.
This enhancement has been theoretically predicted by Eliashberg in Ref.~\cite{Eliashberg} 
and then confirmed experimentally by irradiating a microwave 
to superconducting aluminum films~\cite{Kommers}.
Also, in recent years, 
pump-probe laser measurements of a cuprate, where coherent phonon excitation is induced, 
indicate that some properties in superconducting states can still be observed even at room temperature due to the enhancement~\cite{EnhSC1,EnhSC2}. 
Similar phenomena have also been found in alkali-doped fullerides~\cite{EnhSC3}.
 
The electric field we consider rotates in the $(x,y)$-plane with a frequency $\Omega$ as
\begin{equation}
E_x +iE_y= E e^{i\Omega t}\ ,
\label{intro_rotating_E}
\end{equation}
where $E$ specifies the electric field at $t=0$ and is introduced as a complex value.
While the source is time dependent, 
we can take advantage of the circular polarization~\eqref{intro_rotating_E} 
to construct the solutions in which the time variable separates.
As a result, time independent physical quantities are observed~\cite{Hashimoto:2016ize,Kinoshita:2017uch}.\footnote{Recently, a relativistic fluid with the rotating electric field has been considered in Ref.~\cite{Baumgartner:2018dqi}.}
In particular, we obtain a time independent scalar condensate, which characterizes the phase structure of our Floquet system.

For a linearly polarized electric field, $E_x \propto \cos \Omega t$ and $E_y=0$, 
the superconducting order parameters become time dependent, and it is necessary to address $(1+1)$-dimensional problems~\cite{Li:2013fhw,Natsuume:2013lfa}.\footnote{Time periodic driving of scalar field with an external scalar source $\phi_0 \propto \cos \Omega t$ in AdS/CFT has been studied in Refs.~\cite{Auzzi:2012ca,Auzzi:2013pca,Rangamani:2015sha}.}
In Ref.~\cite{Natsuume:2013lfa}, small perturbations on the normal phase were studied, 
and it was shown that, at least ``near'' the normal phase (in the 
infinitesimal limit of the superconducting order parameter), 
no enhancement happens to the superconductivity.
Ref.~\cite{Li:2013fhw} directly treated the $(1+1)$-dimensional partial differential equations under periodic driving by a linearly polarized electric field, and studied the time evolution and relaxation of superconducting order parameters.

The rotating electric field \eqref{intro_rotating_E} can be derived from the gauge potential
\begin{equation}
A_x+iA_y=Ae^{i\Omega t}\ ,
\label{intro_rotating_A}
\end{equation}
where $A \equiv i E/\Omega$.
Taking a non-rotating limit $\Omega \to 0$ with $A$ being fixed, we obtain a constant gauge potential $A_x+iA_y=A$.
Holographic setups in the presence of such a constant source have been studied as superfluids with a supercurrent~\cite{Basu:2008st,Herzog:2008he,Arean:2010xd,Sonner:2010yx,Arean:2010zw}.
While we assume that we apply a rotating electric field to superconductive materials, the finite-$\Omega$ case of ours can also be interpreted as a superfluid with a rotating current.

Steady state thermodynamics is an important topic in Floquet physics~\cite{Kohn:2014,Lazarides:2014,Lazarides:20142,DAlessio:2014}.
Can we define a ``free energy'' that determines the ``thermodynamical'' stability of the Floquet states?
The AdS/CFT correspondence might help us to understand such a problem.\footnote{%
See also, for example, Ref.~\cite{Nakamura:2012ae} 
for an attempt to construct the steady state thermodynamics 
under a DC electric field in AdS/CFT.}
Using the AdS/CFT consideration, we will examine the difficulty in defining the free energy for nonequilibrium systems which tells us the phase structure.
Following that, we propose that one reasonable definition of the free energy is given by
the Maxwell construction.\footnote{In the presence of a DC electric field, the Maxwell construction is used as the primary option to determine the nonequilibrium phase structure~\cite{Albash:2007bq,Bergman:2008sg}. It can be also used in the case of a magnetic field~\cite{Albash:2007bk,Erdmenger:2007bn}.}

\begin{figure}
\centering 
\includegraphics[scale=0.7]{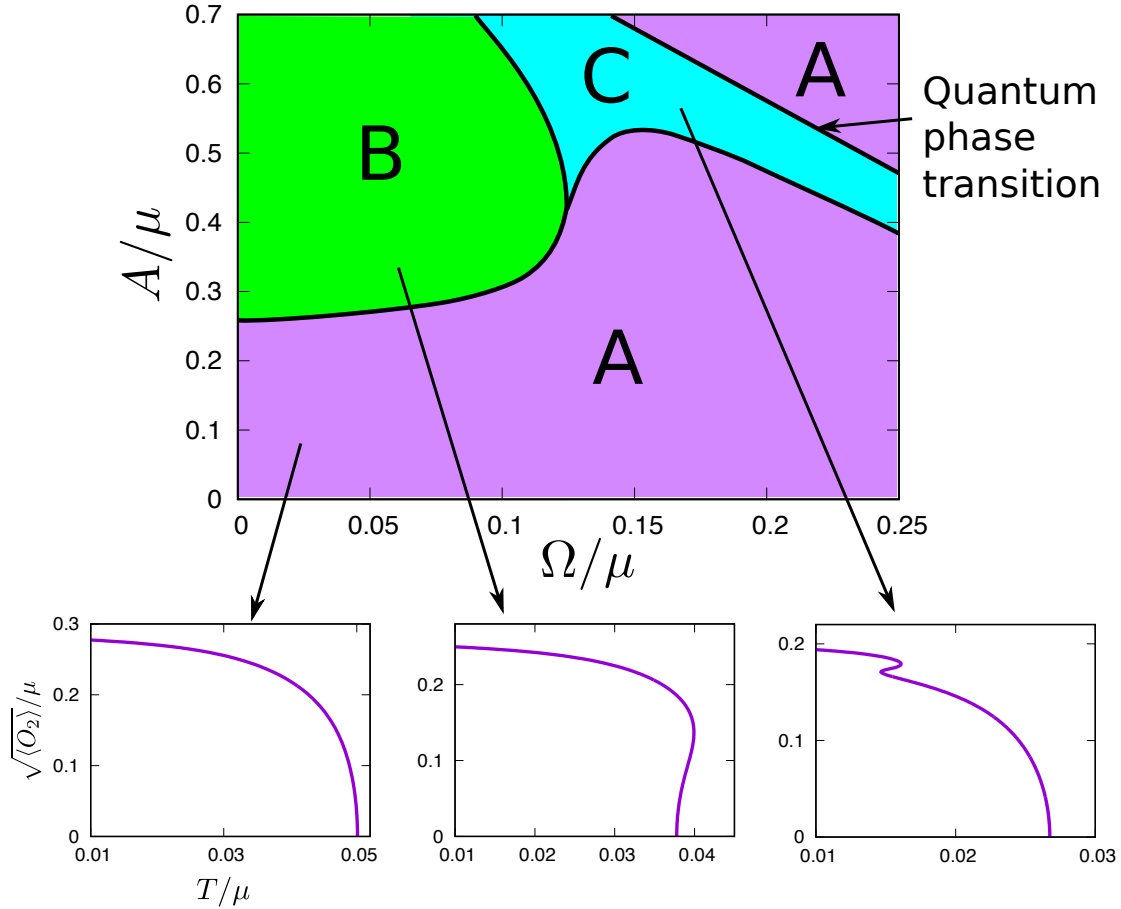}
\caption{
Phase diagram.
The space of $(A/\mu,\Omega/\mu)$ is separated into three regions A, B and C.
In each region, the superconducting order parameter $\langle O_2\rangle$ behaves qualitatively differently as a function of temperature $T$. On the line labeled ``quantum critical transition'', a $T=0$ first order quantum phase transition is expected.
}
\label{phase0}
\end{figure}

Before going to detailed analysis in the following sections, 
we briefly describe the main results of this paper.
The parameters we tune are temperature $T$, chemical potential $\mu$, the 
amplitude of the gauge potential $A$, and frequency of the rotating electric field $\Omega$.\footnote{%
We find it practical to use the gauge potential $A$ as a parameter instead of the electric field $E$.}
While $A$ is treated as a complex value in the main text, in the plots we adjust its phase to zero so that $A$ is real and corresponds to the amplitude without loss of generality.
In this paper, we consider the grand canonical ensemble and measure physical quantities in units of $\mu$.
In Fig.~\ref{phase0}, we provide a summary of phase transition patterns in $(A,\Omega)$-space, where the dependence on $T$ is suppressed.
There are three regions, A, B and C characterized by the behavior of the physical quantities. 
In each region, we find qualitatively different temperature dependence 
of the superconducting order parameter $\langle O_2\rangle$, typical examples of which are shown in the insets in the bottom of the figure.
These represent different phase-transition behaviors in A, B and C.
In region A, the order parameter is monotonic in temperature.
There is a second-order phase transition from normal to superconducting phases.
The order parameter behaves as $\langle O_2 \rangle \propto (1-T/T_c)^{1/2}$ near the critical temperature $T_c$.
In region B, the order parameter becomes multivalued 
around the temperature at which the superconducting phase appears.
It exhibits a first order superconducting phase transition that
the order parameter jumps from zero to a finite value.
The phase structure for the static case $\Omega=0$ has been obtained in Refs.~\cite{Basu:2008st,Herzog:2008he}.
The first order superconducting transition in the presence of 
the time-dependent external field has been theoretically predicted in Refs.~\cite{Ivlev1,Ivlev2}
and experimentally observed in Ref.~\cite{Kommers}.
As $\Omega$ is increased, region B disappears in $\Omega/\mu\gtrsim 0.12$.
Region C is a new phase which appears only in $\Omega\neq 0$: 
The order parameter is multivalued {\textit{only inside}} the superconducting phase.
In this region, we experience two phase transitions as the temperature lowers.
The first is a second order transition from normal to superconducting phases.
The second is a first order transition inside the superconducting phase.
We also find that the latter results in a first order quantum phase transition (at $T=0$).
On the upper-right boundary of the region C, the transition temperature becomes zero.\footnote{%
The boundary of the region C is found by numerical extrapolation.
See appendix~\ref{app:DetailPhase} for details.}
Even if we keep the system at absolute zero, we can realize the first order 
quantum phase transition by controlling $A$ or $\Omega$.
Quantum phase transitions inside superconducting phases have been observed in 
some heavy-fermion metals by using a magnetic field, 
chemical substitution or pressure 
as non-thermal control parameters~\cite{ReviewQCP}.
Our holographic calculations 
indicate that similar phenomena can be caused by the rotating electric field instead of the static control parameters.
We also comment that, in our results, the critical temperature for the appearance of the superconductivity is always lower in the presence of the electric field than its absence. Thus we do not observe any enhancement of superconductivity.

In the rest of the paper, we start by introducing our setup in section~\ref{sec:model}. We show results of the order parameter in section~\ref{sec:vev} and construct the phase diagram in section~\ref{sec:phase}. The current induced by the electric field is studied in section~\ref{sec:other}.

\section{A holographic superconductor with a rotating electric field}
\label{sec:model}

\subsection{Time dependent setup}

We consider a holographic model for superconductivity introduced in Refs.~\cite{Gubser:2008px,Hartnoll:2008vx,Hartnoll:2008kx}:
\begin{equation}
 S=-\int d^4x \sqrt{-g}\left[
\frac{1}{4}F^{\mu\nu}F_{\mu\nu}+|D\Psi|^2-2L^{-2}|\Psi|^2
\right]\ ,
\label{actionHSC}
\end{equation}
where $D_\mu\equiv \nabla_\mu -iA_\mu$ and $L$ is the AdS scale.
The mass of the scalar is set as $m^2=-2/L^2$.
We work in the probe limit where the backreaction to the geometry is absent.
The background geometry is the Schwarzschild-AdS$_4$ solution:
\begin{equation}
 ds^2 = \frac{L^2}{z^2}\left[-f(z)dt^2+\frac{dz^2}{f(z)}+dx^2+dy^2\right]\ ,\quad f(z)=1-\frac{z^3}{z_h^3}\ ,
\label{SchAdS}
\end{equation}
where $z=0$ and $z_h$ correspond to the AdS boundary and event horizon, respectively.
The Hawking temperature is 
\begin{equation}
 T=\frac{3}{4\pi z_h}\ .
\end{equation}
Hereafter, we take units where $L=1$.

We will construct classical solutions of the system~(\ref{actionHSC}) when the rotating electric field \eqref{intro_rotating_E} is applied.
We can assume all variables do not depend on the spatial coordinates of the AdS boundary: $A_\mu=A_\mu(t,z)$ and $\Psi=\Psi(t,z)$.
This is a consistent assumption because of the isometry $\partial_x$ and $\partial_y$ in the background geometry~(\ref{SchAdS}).
Meanwhile, we keep the time dependence to take into account the rotating electric field.
For the convenience of subsequent computations, we introduce a complex notation for the gauge field,
\begin{equation}
 a(t,z)=A_x(t,z)+iA_y(t,z)\ .
\label{compex_a}
\end{equation}
Under the above ansatz, the Lagrangian is explicitly written as
\begin{equation}
\mathcal{L}= 
 \frac{1}{2}F_{tz}^2 
+\frac{1}{2f} |\dot{a}|^2 
-\frac{1}{2}f |a'|^2
+\frac{1}{z^2}\left[
\frac{1}{f} |D_t \Psi|^2 
-f  |D_z \Psi|^2
+\frac{2}{z^2} |\Psi|^2
- |a\Psi|^2\right]\ ,
\label{Lag1}
\end{equation}
where $\dot{}=\partial_t$ and ${}'=\partial_z$.
Near the AdS boundary, the gauge field is expanded as
\begin{equation}
a(t,z)=\mathcal{A}(t)+\mathcal{J}(t) z+\cdots\ .
\end{equation}
In the expansion, $\mathcal{A}(t)$ corresponds to the external gauge potential in the dual field theory. 
The external electric field is given by
\begin{equation}
 \mathcal{E}(t)=-\dot{\mathcal{A}}(t)\ .
\end{equation}
The second leading term $\mathcal{J}(t)$ corresponds to its response: electric current.
Since we use the complex variable as in Eq.~(\ref{compex_a}), 
$\mathcal{A}(t)$ and $\mathcal{J}(t)$ also are complex. 
The real and imaginary parts of each variable are its $x$- and $y$-components.

In terms of the complex field, the external rotating source \eqref{intro_rotating_E} and \eqref{intro_rotating_A} is written as\footnote{Recently, Ref.~\cite{Biasi:2017kkn} studied Floquet dynamics in a holographic model with a complex scalar with a rotating phase.}
\begin{equation}
\mathcal{A}(t)=A e^{i\Omega t}\ ,\quad \mathcal{E}(t)=E e^{i\Omega t}\ ,\quad A=i\frac{E}{\Omega}\ .
\label{rotE}
\end{equation}
To consider the gauge field with the condition~(\ref{rotE}),
it is convenient to factor out the phase and introduce a new field $b(t,z)$ as
\begin{equation}
a(t,z)=e^{i\Omega t} b(t,z)\ .
\label{aasb}
\end{equation}
The boundary condition for $b(t,z)$ is simply given by $b|_{z=0}=A$.
Written in the new variable $b$, the Lagrangian~(\ref{Lag1}) becomes
\begin{multline}
\mathcal{L}=
 \frac{1}{2}F_{tz}^2 
+ \frac{1}{2f} |(\partial_t+i\Omega)b|^2 
-\frac{1}{2}f |b'|^2\\
+\frac{1}{z^2}\left[
\frac{1}{f} |D_t \Psi|^2 
-f  |D_z \Psi|^2
+\frac{2}{z^2} |\Psi|^2
-|b\Psi|^2\right]\ .
\end{multline}
Remarkably, this Lagrangian does not depend on $t$ explicitly.\footnote{%
Although we can also take into account the rotating phase for the complex scalar field as
$\Psi(t,z)=e^{i\Omega' t}\Phi(t,z)$, 
we have gauged out this phase factor and set $\Omega'=0$.}
Therefore, we can consistently assume that the variables ($A_t, A_z, b, \Psi$) do not depend on $t$.
Then, the Lagrangian becomes
\begin{multline}
\mathcal{L}=
- \frac{1}{2}A_t'{}^2 
- \frac{\Omega^2}{2f} |b|^2 
+\frac{1}{2}f |b'|^2\\
+\frac{1}{z^2}\left[
-\frac{1}{f} A_t^2|\Psi|^2 
+f |(\partial_z - i A_z) \Psi|^2
-\frac{2}{z^2} |\Psi|^2
+ |b \Psi|^2
\right]\ .
\label{1dLag}
\end{multline}
While the external electric field is time dependent~(\ref{rotE}), 
the system reduces to a one-dimensional problem in the $z$-direction.
As mentioned in Refs.~\cite{Hashimoto:2016ize,Kinoshita:2017uch},
this is a special property of the rotating electric field.
If a linearly polarized electric field is considered such as 
$\mathcal{E}_x\propto \cos \Omega t$ and $\mathcal{E}_y=0$, one needs to address ($1+1$)-dimensional problems~\cite{Li:2013fhw}.

An external electric field could heat up the system.
For a physical application, therefore, the backreaction of the electric field to the background metric may be important.
But here we work in the probe limit, where the temperature of the background is not changed by the electric field and the Schwarzschild solution acts as a thermal bath, realizing steady solutions.
We would mainly like to extract constructive lessons from this ideal setup.
The problem of the backreaction will be addressed elsewhere.

From the Lagrangian~(\ref{1dLag}), we obtain time independent equations of motion. 
Using the gauge degrees of freedom, we adopt a gauge in which $A_z(z)=0$.
Taking the variation of the above Lagrangian by $(A_t,A_z,b,\Psi)$ and imposing the gauge condition, we obtain
\begin{align}
&\Psi''= \left(\frac{2}{z}-\frac{f'}{f}\right)\Psi'
+\frac{1}{f}\left(|b|^2- \frac{A_t^2}{f}-\frac{2}{z^2}\right)\Psi\ ,\label{psieq}\\
&b''=-\frac{f'}{f}b'+\frac{1}{f}\left(-\frac{\Omega^2}{f}+\frac{2|\Psi|^2}{z^2}\right)b\ ,\label{beq}\\
&A_t''=\frac{2|\Psi|^2}{z^2 f}A_t\ ,\label{Ateq}\\
&\Psi^\ast\Psi'-\Psi\Psi'{}^\ast=0\ .\label{const}
\end{align}
The last equation is the constraint obtained from the variation of the Lagrangian with respect to $A_z$.
It implies that the phase of $\Psi$ is constant, and by a gauge transformation we can set $\Psi$ to be real without loss of generality.\footnote{That is, we use the unitary gauge. In general, a gauge invariant combination $A+d\varphi$ is used for the gauge potential for the variation where $\varphi$ is the Nambu-Goldstone boson $\Psi \to \Psi e^{i \varphi}$.}

Solving the equations of motion near the AdS boundary, 
we obtain the asymptotic expansions of the bulk fields, 
\begin{equation}
b(z)=A+J z + \cdots\ ,\quad
A_t(z)=\mu-\rho z+\cdots \ ,\quad
\Psi(z)=\psi_1 z + \psi_2 z^2+\cdots \ ,
\label{bstatic_asym}
\end{equation}
where $\mu$ and $\rho$ correspond to the chemical potential and charge density.
The complex constant $J$ appears in the rotating electric current as
\begin{equation}
\mathcal{J}(t)=Je^{i\Omega t}\ .
\label{rotJ}
\end{equation}
As the boundary condition of $\Psi$, one can impose $\psi_1=0$ or $\psi_2=0$~\cite{Klebanov:1999tb,Hartnoll:2008vx,Hartnoll:2008kx}.
Here we impose $\psi_1=0$ and take 
the scalar condensate as a dimension-two operator:
\begin{equation}
\langle O_2\rangle = \sqrt{2}\psi_2\ .
\end{equation}
We defined the response functions from subleading terms in the asymptotic expansion of the bulk fields. This is a quick way 
to read out the response functions. Actually, in Refs.\cite{Herzog:2002pc,Skenderis:2008dg}, 
it is shown that subleading terms in the asymptotic expansion can be
regarded as one-point functions of operators in the boundary theory even for the real-time AdS/CFT.

The reduced Lagrangian~(\ref{1dLag}) is invariant under $b\to be^{i\theta}$ with a constant $\theta$.
The Noether charge corresponding to this symmetry gives
\begin{equation}
q=\frac{i\Omega}{2}f(z)(b'b^\ast-bb'{}^\ast)\ .
\label{qdef}
\end{equation}
This is conserved in the $z$-direction: $\partial_z q=0$.
To see its physical meaning, we substitute (\ref{bstatic_asym}) into (\ref{qdef}) and take $z\to 0$.
We then obtain
\begin{equation}
 q=\textrm{Re}(EJ^\ast)=\vec{\mathcal{E}}\cdot\vec{\mathcal{J}}\ ,
\end{equation}
where $\vec{\mathcal{E}}=(\textrm{Re}\mathcal{E}, \textrm{Im}\mathcal{E})=(\mathcal{E}_x,\mathcal{E}_y)$ and so is $\mathcal{J}$.
This is nothing but the Joule heating generated by the rotating electric field.

This system has a scaling symmetry: 
\begin{equation}
\begin{split}
&\psi_1\to \alpha \psi_1\ ,\quad
 \psi_2\to \alpha^2 \psi_2\ ,\quad
 \mu\to \alpha \mu\ ,\quad 
 \rho \to \alpha^2 \rho\ ,\\
&\Omega\to \alpha \Omega\ ,\quad
 A\to \alpha A\ ,\quad
 J\to \alpha^2 J\ , \quad
 T\to \alpha T\ ,
\end{split}
\label{scalingsym}
\end{equation}
for $\alpha\neq 0$.
In this paper, we work with a fixed chemical potential and non-dimensionalize the physical quantities by $\mu$.
This corresponds to working in the grand canonical ensemble in the dual field theory.

To solve the equations of motion, we also need the boundary condition at the black hole horizon.
A regular and ingoing-wave solution near the event horizon is given by
\begin{equation}
\Psi\simeq \Psi_H\ ,\quad b\simeq b_H e^{i \Omega r_\ast}\ ,\quad
 A_t\simeq A_H (z_h-z)\ ,
\label{horbc}
\end{equation}
where we introduced a tortoise coordinate,
\begin{equation}
 r_\ast=-\int^z_0 \frac{dz'}{f(z')}\ .
\end{equation}
Note that, when $\Psi\neq 0$, a constant term for $A_t$ is not allowed 
from the regularity of the future event horizon~\cite{Gubser:2008px}. 
(An observer falling into the future event horizon with 4-velocity $u^\mu$ feels infinite energy 
$E_\textrm{obs}=T_{\mu\nu}u^\mu u^\nu$ unless Eq.~(\ref{horbc}) is satisfied.)

\subsection{Normal phase solution}
\label{normalsol}

In the normal phase $\Psi=0$, the equations of motion can be explicitly solved by
\begin{equation}
b= A e^{i\Omega r_\ast}\ ,\quad A_t=\mu \left(1-\frac{z}{z_h}\right)\ .
\label{nomal_sol}
\end{equation}
Near the AdS boundary, $b(z)$ is expanded as
\begin{equation}
b(z)=A(1-i\Omega z)+\cdots\ .
\label{nomal_sol_expand}
\end{equation}
Comparing Eqs.~(\ref{nomal_sol}) and (\ref{nomal_sol_expand}) with Eq.~(\ref{bstatic_asym}), we obtain the electric current and charge density in the normal phase as 
\begin{equation}
 J=-i\Omega A = E\ , \qquad \rho=\frac{\mu}{z_h}=\frac{4\pi\mu T}{3}\ .
\label{nomalPhysq}
\end{equation}
The electric current depends only on the electric field and is always parallel to the electric field.
The Joule heating is $q=|E|^2$ in the normal phase.

\subsection{Condensed phase solutions}
\label{supersol}

In the condensed phase $\Psi\neq 0$, we need to solve the bulk equations (\ref{psieq}-\ref{const}) numerically.
Details of our numerical implementation are explained in appendix~\ref{Numerics}.
On the horizon, we impose \eqref{horbc}. At the AdS boundary, we specify $\mu$, $A$ and $\psi_1=0$ as the theory's boundary sources.
The equations also contain $T$ and $\Omega$ as parameters.
Other boundary quantities $\rho, \, J$ and $\psi_2$ are read off from the asymptotic form \eqref{bstatic_asym}. 
These are assembled into dimensionless quantities in units of $\mu$:
The field theory parameters we tune are $(T/\mu, A/\mu, \Omega/\mu)$, and we obtain $\rho/\mu^2$, $\langle O_2 \rangle/\mu^2$ and $J/\mu^2$ as physical output. 

\subsection{Perturbative analysis for small scalar condensate}
\label{perturbation}

The onset of the scalar's condensation can be analyzed by using a small perturbation of the scalar field in the normal phase. From Eq.~(\ref{psieq}), 
the equation for the perturbation is given by
\begin{equation}
\Psi''= \left(\frac{2}{z}-\frac{f'}{f}\right)\Psi'
+\frac{1}{f}\left(|A|^2- \frac{\mu^2}{f}\left(1-\frac{z}{z_h} \right)^2-\frac{2}{z^2}\right)\Psi\ .
\label{1steq}
\end{equation}
We solve this equation by imposing $\psi_1=0$ at the AdS boundary and regularity on the horizon
and searching for the value of $T/\mu$ at which a nontrivial $\Psi$ becomes a solution.
This $T/\mu=T_\textrm{branch}/\mu$ gives the critical temperature when $\langle O_2 \rangle$ starts to become nonzero and the superconducting solution branches from the trivial solution.
We call this point the branching point or branching temperature.
The branching temperature can be different from the temperature of the phase transition, specifically when there is a first order superconducting phase transition, 
as we will see in the next section.
In Fig.~\ref{fig.Tb}, $T_\textrm{branch}/\mu$ is shown as a function of $A/\mu$.
As seen in \eqref{1steq}, $T_\textrm{branch}$ depends only on $A/\mu$ and not on $\Omega/\mu$ independently.
Thus the branching temperature agrees with the $\Omega=0$ result \cite{Basu:2008st,Herzog:2008he} in any $\Omega$.
It is a decreasing function of $A$ and the maximum is $T_\textrm{branch}/\mu=0.0587$ at $A=0$.
In Ref.~\cite{Natsuume:2013lfa}, 
the branching temperature has been studied for a linearly polarized periodic electric field and 
it is shown that the temperature is always decreased by the effect of the electric field 
at least in sufficiently small or high frequency limits.
We obtain the same consequence in the case of the rotating electric field with general $\Omega$. 
We will also see that the phase transition temperature, which can be different from $T_\textrm{branch}$, 
is also lower than $T_\textrm{branch}/\mu=0.0587$.

\section{Superconducting order parameter}
\label{sec:vev}

\begin{figure}
  \centering
  \subfigure[$A/\mu=0.25$]
 {\includegraphics[scale=0.45]{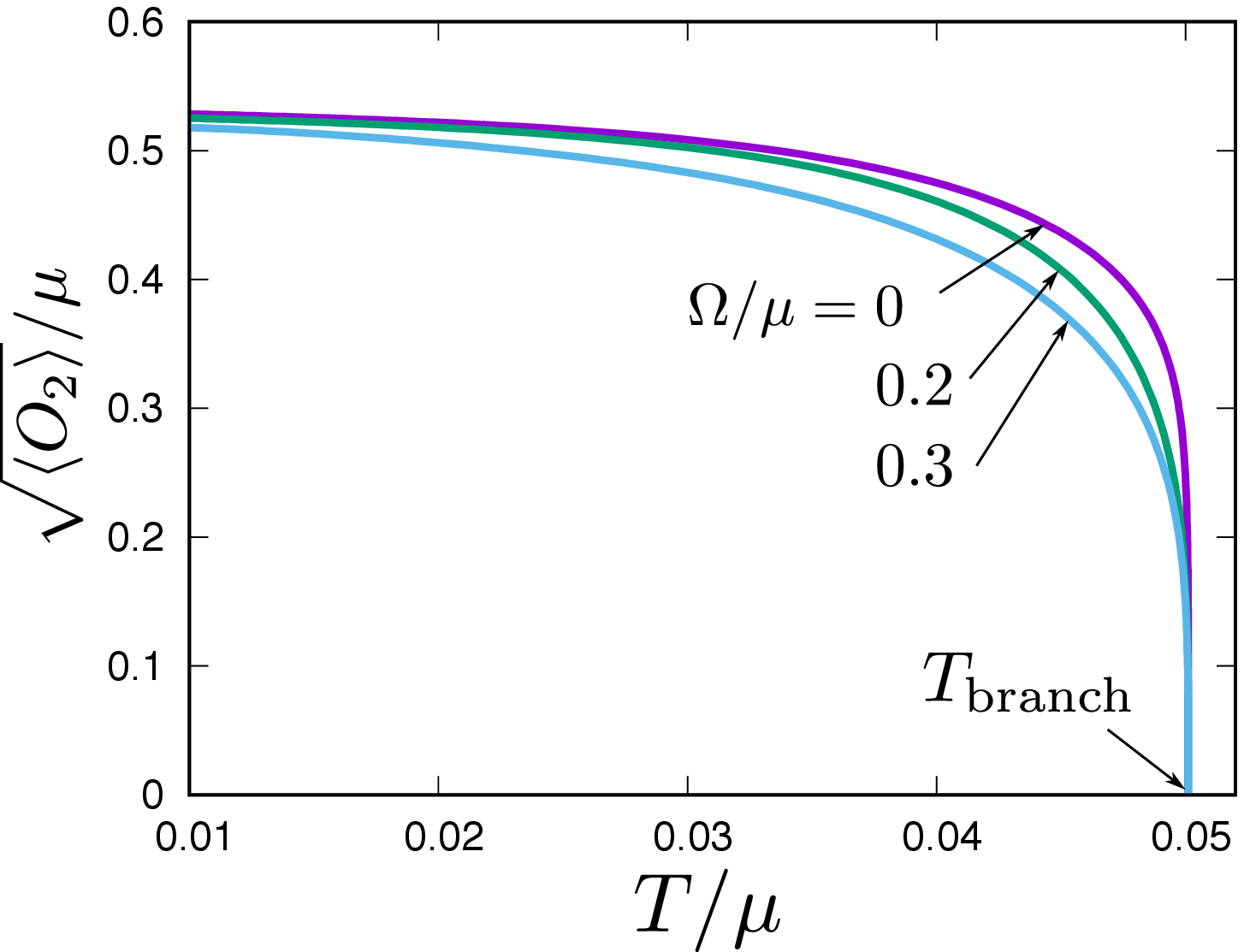}\label{O2_025}
  }
\subfigure[$A/\mu=0.4$]
 {\includegraphics[scale=0.45]{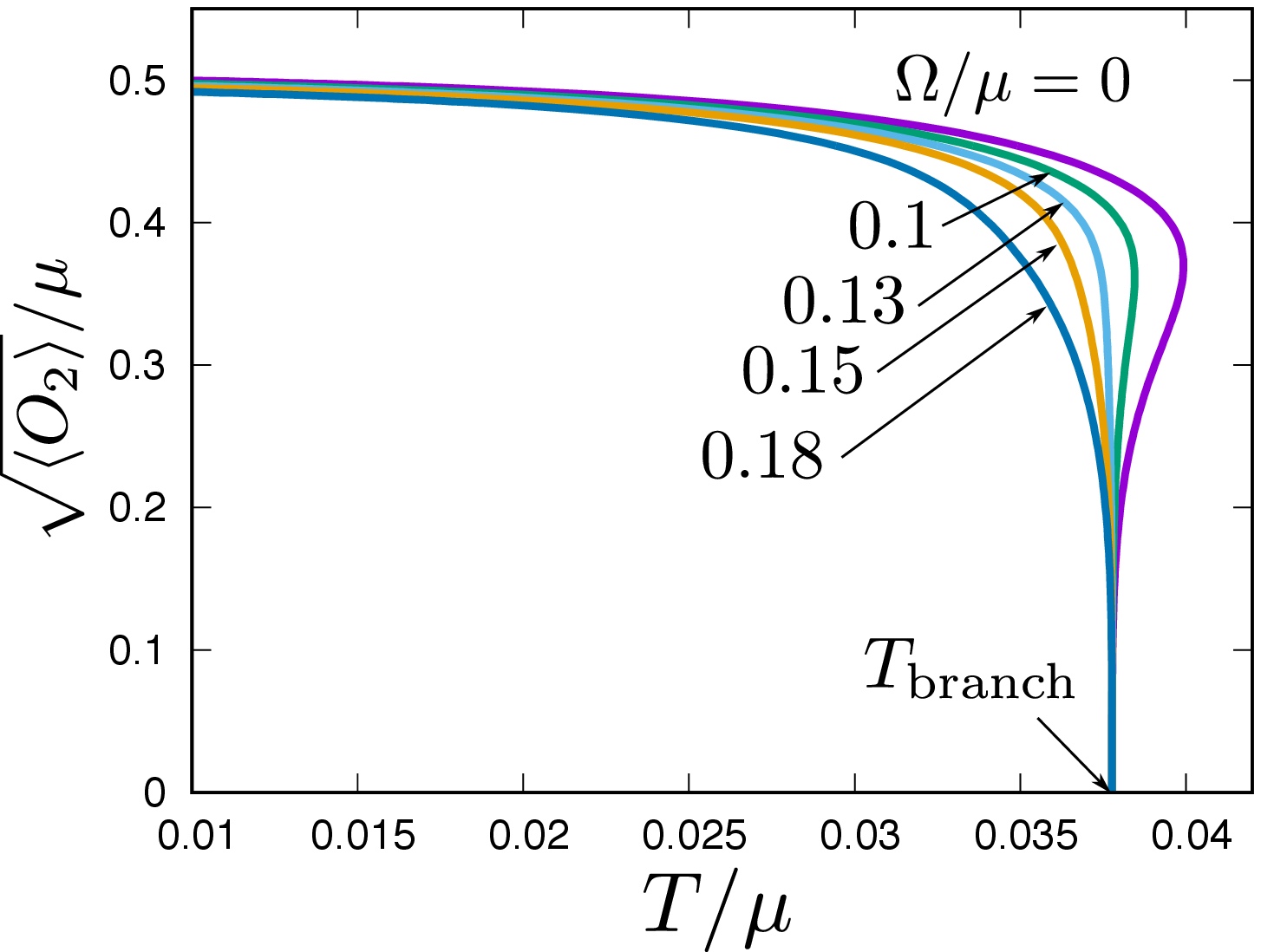}\label{O2_040}
  }
  \subfigure[$A/\mu=0.53$]
 {\includegraphics[scale=0.45]{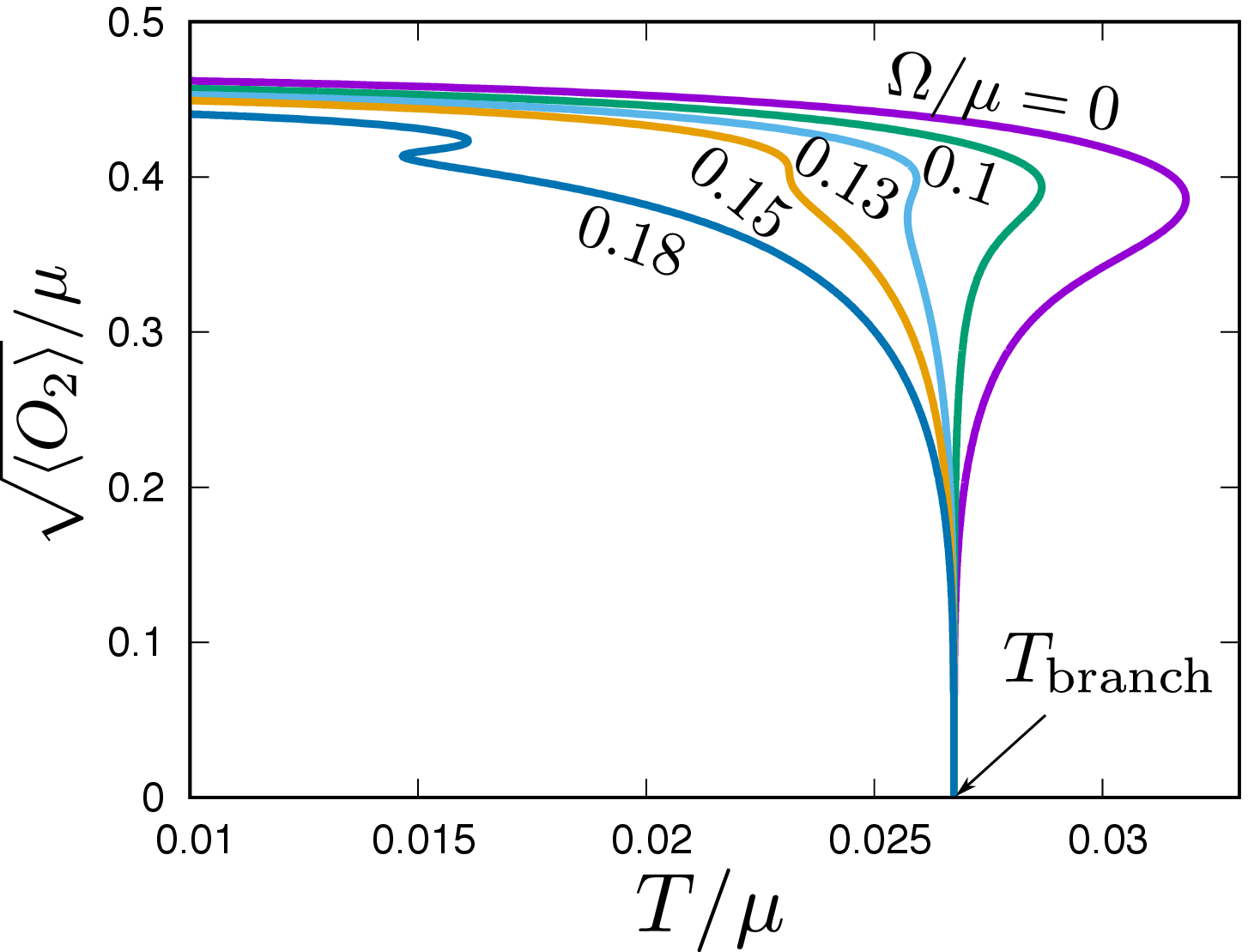}\label{O2_053}
  }
 \subfigure[Branching temperature]
 {\includegraphics[scale=0.45]{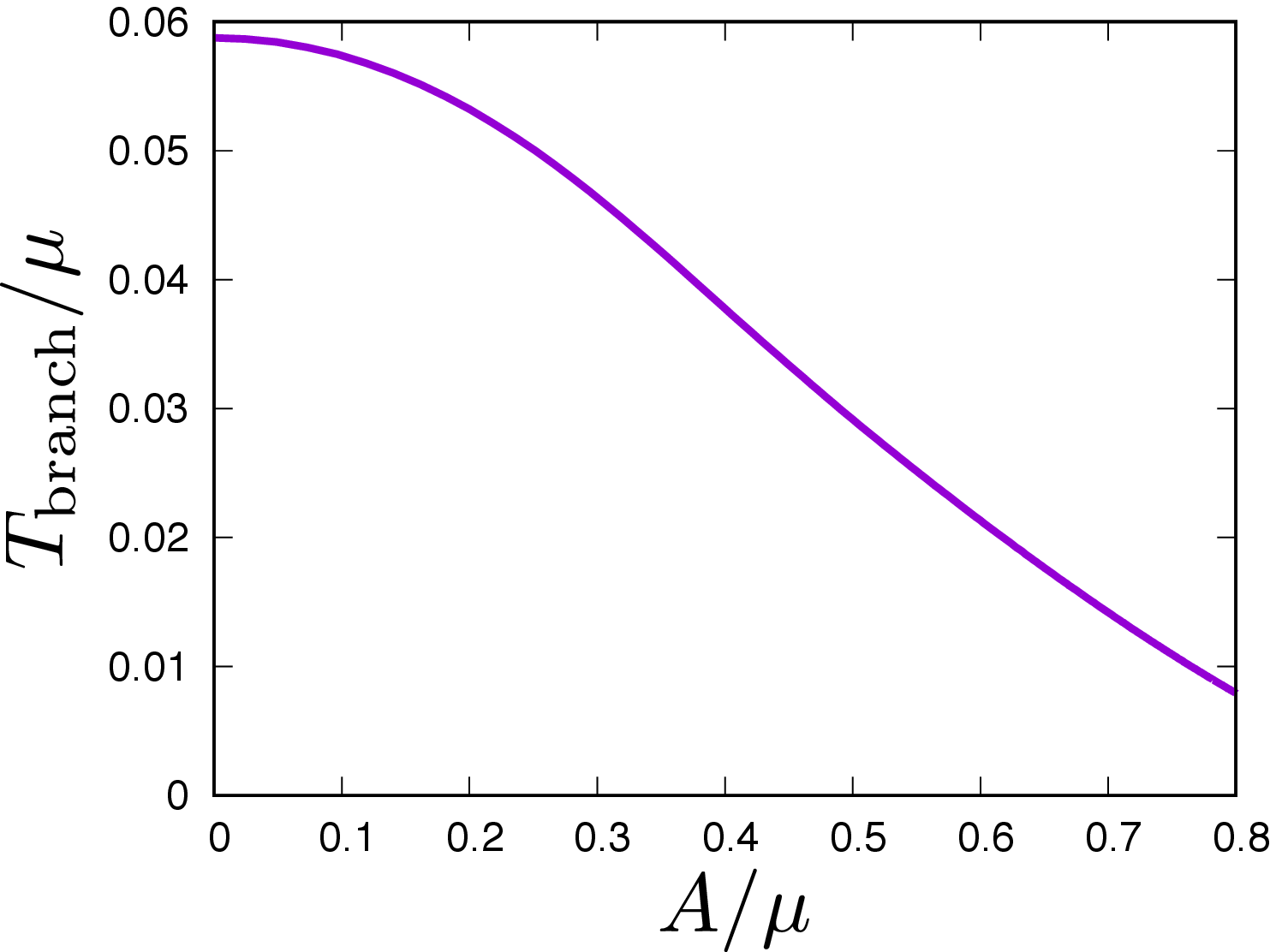}\label{fig.Tb}
  }
 \caption{
Panels (a-c): Temperature dependence of the superconducting order parameter $\sqrt{\langle O_2\rangle}/\mu$ 
at $A/\mu=0.25,0.4,0.53$.
Panel (d): The $A/\mu$-dependence of the branching temperature where $\langle O_2\rangle$ starts to become nonzero. This does not depend on $\Omega$ for fixed $A$.
\label{O2}
}
\end{figure}

In the following, we show results.
Firstly, we look at the superconducting order parameter $\langle O_2\rangle$.
Fig.~\ref{O2} shows $\langle O_2\rangle$ as a function of $T$ at different $A$ and $\Omega$.
As mentioned before, we non-dimensionalize the physical quantities by the chemical potential $\mu$.
Although the electric field $E$ might be a physically intuitive parameter in the boundary theory, 
we find it more practical to use the amplitude of the gauge potential $A=iE/\Omega$ as a parameter.
For example, in the static limit $\Omega\to 0$,
the gauge potential diverges if $E$ is fixed, while a regular limit can be taken if we fix $A$.
From $A$, one can easily obtain the value of the electric field as $E/\mu^2=-i A/\mu \times \Omega/\mu$.
We observe that the condensed solution ($\langle O_2\rangle\neq 0$) branches from 
the normal phase ($\langle O_2\rangle= 0$) at $T=T_\textrm{branch}$. 
Fig.~\ref{fig.Tb} shows the $A$-dependence of the branching temperature studied in section~\ref{perturbation}.
In the following paragraphs, we explain each panel in detail.

In panel (a), we fix the amplitude of the gauge potential to a relatively small value, $A/\mu=0.25$, and 
vary the frequency of the rotating electric field as $\Omega/\mu=0,0.2,0.3$.
These results are qualitatively similar to the zero electric field case originally studied in Ref.~\cite{Hartnoll:2008vx}:
The scalar condensate increases sharply as $\langle O_2\rangle \sim \sqrt{1-T/T_c}$ near the critical temperature 
and is a single valued function in all $T$. 
By the effect of the rotating electric field, the scalar condensate decreases as the frequency increases.

In panel (b), we take 
an intermediate value for the gauge potential, $A/\mu=0.4$.
The frequency is varied as $\Omega/\mu=0,0.1,0.13,0.15,0.18$.
In the static limit $\Omega=0$, we reproduced the results in Ref.~\cite{Herzog:2008he}:
The curve for $\langle O_2 \rangle$ bends to the right near the branching point and becomes multivalued as a function of $T$. 
This indicates that there is a first order phase transition at $T>T_\textrm{branch}$. 
As we increase $\Omega$, the curve is ``pushed'' to the left and the multivaluedness 
is eventually resolved around $\Omega/\mu\simeq 0.13$.

In panel (c), the gauge potential is fixed to a relatively large value, $A/\mu=0.53$.
The frequencies are $\Omega/\mu=0,0.1,0.13,0.15,0.18$.
At $\Omega=0$, we again find a multivalued profile. 
Although the curve moves to the left as $\Omega$ increases, 
the multivaluedness remains even in $T<T_\textrm{branch}$ 
and $\langle O_2(T) \rangle$ develops an ``inverse S-shape''. (See the curve for $\Omega/\mu=0.13$.)
This indicates that there is a first order phase transition inside the superconducting phase.
If we increase $\Omega$ further, the multivaluedness is resolved once around $\Omega/\mu\sim 0.15$,
but again it becomes inverse S-shaped at $\Omega/\mu\sim 0.18$.
From panel (c), we observe that 
the scalar condensate can become multivalued in a complicated way, indicating a rich phase structure in the Floquet holographic superconductor as we study in the next section.

\section{Phase structure}
\label{sec:phase}

\subsection{Free energy and Maxwell construction}
\label{Maxwell}

\subsubsection{Equilibrium case: $\Omega=0$}

Firstly, let us focus on the static case $\Omega=0$ studied in Ref.~\cite{Herzog:2008he}.
The Lorentzian renormalized on-shell action takes the form\footnote{%
In this section, for the visibility of the relations between the sources and responses, we use the vector notation instead of the complex one as
$\vec{J}=(\textrm{Re}J,\textrm{Im}J)$, etc.}
\begin{equation}
S_\textrm{ren}=\int dt d^2x \mathcal{L}\ ,\quad
\mathcal{L}\equiv \frac{1}{2}\left(\mu\rho+\vec{A}\cdot \vec{J}\right) + \int^{z_h}_0 \frac{dz}{z^2}\left(|b|^2-\frac{A_t^2}{f}\right)\Psi^2 \ ,
\label{Sreg0}
\end{equation}
where $\psi_1=0$ is used. 
In thermal equilibrium, we go to the Euclidean signature by $\int dt \to -i\int^\beta_0 d\tau$.
The Euclidean on-shell action is then written as $S_E = -\int d\tau d^2x \mathcal{L}$.
The free energy density is given by the on-shell action as $F_{\Omega=0}=TS_E/V=-\mathcal{L}$ where $V=\int d^2x$.
Thus we obtain
\begin{equation}
 F_{\Omega=0} = - \frac{1}{2}\left(\mu\rho+\vec{A}\cdot \vec{J}\right) - \int^{z_h}_0 \frac{dz}{z^2}\left(|b|^2-\frac{A_t^2}{f}\right)\Psi^2 \ .
\label{F0def}
\end{equation}

This is a satisfactory definition of the free energy in the sense that 
it gives a generating function of the responses $\rho$ and $\vec{J}$ as
\begin{equation}
\frac{\partial F_{\Omega=0}}{\partial \mu}=-\rho\ ,\quad 
\frac{\partial F_{\Omega=0}}{\partial \vec{A}}=-\vec{J} \ .
\label{diffF0}
\end{equation}
These relations are obtained by taking the variation of the on-shell action.
From the Euler-Lagrange equation, the variation of the on-shell action becomes the boundary term solely from the AdS boundary,
\begin{equation}
\delta S_\textrm{ren} = \int dt d^2x \left(\rho \delta \mu + \vec{J}\cdot \delta \vec{A}\right)\ ,
\label{s_ren_variation_0}
\end{equation}
where we ignore the terms proportional to $\delta \psi_1$ for simplicity because we fix $\psi_1=0$.
The case including $\delta \psi_1$ is discussed in appendix~\ref{app:scalar}.
From~(\ref{s_ren_variation_0}), we have the total derivative of the free energy as
\begin{equation}
dF_{\Omega=0} = -\rho d\mu  - \vec{J}\cdot d \vec{A} \ .
\label{maxcon_dF}
\end{equation}
This form makes it manifest that we work in a grand canonical ensemble. 

The integrability of $dF_{\Omega=0}$ requires that the following conditions be satisfied:
\begin{equation}
\frac{\partial \rho}{\partial \vec{A}}=\frac{\partial \vec{J}}{\partial \mu}\ ,\quad
\frac{\partial}{\partial \vec{A}}\times \vec{J}=0\ .
\label{maxcon_integ}
\end{equation}
We checked that this is indeed fulfilled when $\Omega=0$. 

Once we know the responses $\rho$ and $\vec{J}$ as functions of the sources $\mu$ and $\vec{A}$, it is guaranteed by (\ref{maxcon_integ}) that
the free energy $F_{\Omega=0}$ is constructed by integrating \eqref{maxcon_dF}.
This is the Maxwell construction.
The result is equivalent to (\ref{F0def}) up to the integration constant.
Armed with the clearly defined free energy, we can identify the phase transition in $\Omega=0$ without ambiguity.

\subsubsection{Nonequilibrium case: $\Omega\neq 0$}
\label{NoneqF}

Can the definition of the free energy be naively extended to the nonequilibrium case, $\Omega\neq 0$?
Simply repeating the calculations in Ref.~\cite{Herzog:2008he}, 
we find that the Lorenzian renormalized on-shell action becomes
exactly the same expression as Eq.~(\ref{Sreg0}) even in $\Omega\neq 0$.
Note that the on-shell Lagrangian does not depend on time even though the gauge potential is time dependent,
and this might have implied a naive definition of the free energy: $F_\textrm{naive} \equiv-\mathcal{L}$.\footnote{%
For $\Omega\neq 0$, $\vec{A}$ and $\vec{J}$  express the vector potential and electric current at $t=0$. 
The rotating vector potential and current are related to them by 
$\vec{\mathcal{A}}(t)=R(\Omega t)\vec{A}$ and $\vec{\mathcal{J}}(t)=R(\Omega t)\vec{J}$ 
where $R(\theta)$ is the rotation matrix in the $(x,y)$-plane. See (\ref{rotE}) and (\ref{rotJ}) for the complex notation.
}
However, it is unsatisfactory:
This is not a generating function of the responses, 
$\partial F_\textrm{naive}/\partial \mu\neq -\rho$ 
nor $\partial F_\textrm{naive}/\partial \vec{A}\neq -\vec{J}$. 
We checked them by using numerical results.

The naive definition $F_\textrm{naive}$ indeed has a tension with 
the holographically successful derivation of the response functions~\cite{Son:2002sd}.
To see why this naive $F_\textrm{naive}$ cannot generate the response functions, 
we consider the variation of the on-shell action when $\Omega\neq 0$.
It contains the boundary terms from both the AdS boundary and horizon as
\begin{equation}
\begin{split}
\delta S_\mathrm{reg} &= \delta' s_\infty + \delta' s_h\ ,\\
\delta's_\infty &= \int dt d^2x \left[\rho \delta \mu + \vec{\mathcal{J}}(t)\cdot \delta \vec{\mathcal{A}}(t)\right]\ ,\\
\delta's_h &= -\int dt d^2x\, f\,\textrm{Re}(a'^\ast \delta a)|_{z=z_h}\ ,
\end{split}
\label{s_ren_variation}
\end{equation}
where we use a notation $\delta'$ to indicate the decomposition of $\delta S_\mathrm{reg}$ into the contributions from the horizon and AdS boundary.
However, there is no guarantee that their integrations $s_\infty$ and $s_h$ exist separately.
Actually, we will see that they do not exist in general.
As in Eq.~(\ref{horbc}), the complex gauge field behaves as $a\sim e^{i\Omega(t+r_\ast)}$ near the horizon, and this gives 
a finite $\delta's_h$ when $\Omega\neq 0$.
We obtain
\begin{align}
\delta' s_\infty &=  \int dt d^2x \left[\rho \delta  \mu + \vec{J}\cdot \delta \vec{A} - (q/\Omega) \delta(\Omega t) \right] \ , \\
\delta' s_h &=   \int dt d^2x \left[ \Omega \, \mathrm{Im} (b^\ast \delta b)_{z=z_h} + (q/\Omega) \delta(\Omega t) \right] \ .
\end{align}
The last terms in the above equations for $\delta's_\infty$ and $\delta's_h$ correspond to the conserved flux injected from the boundary and dumped into the horizon, and they cancel.
Thus \eqref{s_ren_variation} becomes
\begin{equation}
\delta S_\mathrm{reg} = \int dt d^2x \left[\rho \delta  \mu + \vec{J}\cdot \delta \vec{A}
+ \Omega \, \mathrm{Im} (b^\ast \delta b)_{z=z_h} \right] \ .
\end{equation}
Note that the last term is absent if $\Omega=0$, and the ingoing wave solution is responsible for that term.
The variation of the naive free energy is hence of the form
\begin{equation}
 d F_\textrm{naive}= -\rho d \mu - \vec{J}\cdot d \vec{A}+
\textrm{horizon contribution}.
\end{equation} 
Because of the nontrivial piece from the horizon, the naive free energy defined from the total onshell action cannot be a generating function for the boundary responses.
This observation is related to the recipe of the AdS/CFT correspondence: We need to use the variation from the AdS boundary and discard that from the horizon in order to obtain the correct responses~\cite{Son:2002sd}.
The boundary responses are generated by $\delta's_\infty$.
That is, the response functions appearing in Eq.(\ref{bstatic_asym}) are obtained from the boundary ``variation'' $\delta' s_\infty$ as
$\delta's_\infty/\delta \mu=\rho$ and 
$\delta's_\infty/\delta \vec{\mathcal{A}}(t)=\vec{\mathcal{J}}(t)$.

\begin{figure}
\centering 
\includegraphics[scale=0.5]{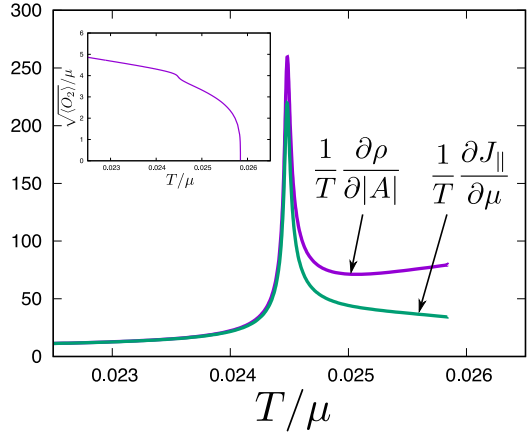}
\caption{
Violation of the integrability condition $\partial \rho/\partial |\vec{A}|=\partial J_\parallel/\partial \mu$
evaluated when $A/T=20.94$ and $\Omega/T=6.283$. 
The chemical potential $\mu$ is varied for a fixed temperature $T$.
The inset shows the scalar condensate for these parameters. 
}
\label{fig:integ_vio}
\end{figure}

This observation would suggest that we employ the Maxwell construction to reasonably define a free energy from the boundary responses
instead of the on-shell action.
This shares the idea of the ``steady state thermodynamics''. (For example, see Ref.~\cite{SST}.)
However, the existence of integrable $dF$ is still problematic:
Unlike the $\Omega=0$ case, 
no function $F$ satisfying at least $dF= -\rho d \mu - \vec{J}\cdot d \vec{A}$ exists.
In fact, \eqref{maxcon_integ} is no longer satisfied once $\Omega \neq 0$.\footnote{
The trajectory dependence of the thermodynamical potential 
in the parameter space of a nonequilibrium system has also been reported 
in Ref.~\cite{Sagawa_Hayakawa}.
}
This can be analytically demonstrated for the normal phase solution~(\ref{nomal_sol}).
From Eq.~(\ref{nomalPhysq}), one can check the violation of the integrability condition: 
We instead obtain $\partial/\partial \vec{A}\times \vec{J}=-2\Omega$.
Also, for the superconducting phase, we numerically test the violation of the integrability condition as shown in Fig.~\ref{fig:integ_vio}.
Since the charge density only depends on $|\vec{A}|$ and $\mu$, 
the first equation of the integrability conditions~(\ref{maxcon_integ}) implies
$\partial \rho/\partial |\vec{A}| =\partial J_\parallel /\partial \mu$ 
where $J_\parallel \equiv \vec{J} \cdot \vec{A}/|\vec{A}|$ is the projection onto the direction of $\vec{A}$.
The plots explicitly show the violation of the integrability condition.
Notably, the violation is significantly small in small $T/\mu$ where $q$ is quite suppressed as we will see later.
This might conversely imply that the system can have a well-posed free energy \eqref{maxcon_dF} 
when the dissipation of the energy is sufficiently small.

While we lack an unambiguous definition of a thermodynamic free energy,
we will still be able to make use of the Maxwell construction to quantify the phase transition in $\Omega \neq 0$.
To evaluate the phase transition in the presence of nonzero $\Omega$, we employ 
\begin{equation}
F \equiv -\int \rho d\mu
\label{Fintrhodmu}
\end{equation}
at fixed $(T, \, A, \, \Omega)$, and use this as the free energy that determines the phase.
Although this definition has an ambiguity to add a free function of $(T,A,\Omega)$ for the integration constant,
it is irrelevant to identifying the phase transition because the difference of the free energies between two states is used.
As discussed in appendix~\ref{app:scalar}, the choice in \eqref{Fintrhodmu} is not the unique one, especially quantitatively, but we find that the $F$ constructed in this way shows the expected behavior for a free energy. (See appendix~\ref{app:DetailPhase}.)
A first order phase transition is expected to be located at some point where the condensate is multivalued, and we can obtain a reasonable estimate out of \eqref{Fintrhodmu}.

\subsection{Phase diagram and transition temperature}

\begin{figure}
\centering 
\includegraphics[scale=0.7]{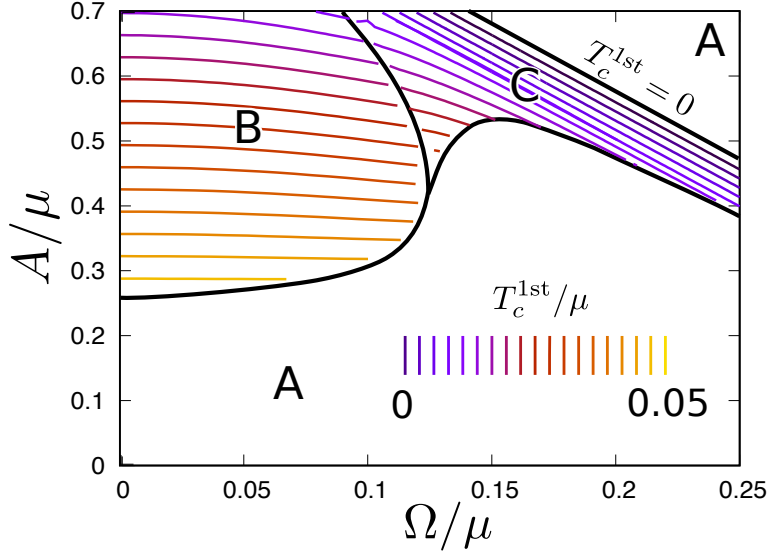}
\caption{
Contour plot of the first order transition temperature $T_c^\textrm{1st}$.
On the upper-right boundary of the region C, the transition temperature reaches zero.
In region A, there is no first order transition. 
For the second order transition temperature, see Fig.~\ref{fig.Tb}.
}
\label{phaseTc}
\end{figure}

Using the Maxwell construction explained above, 
we determine the critical values of $\mu/T$ at which phase transitions occur for fixed $(A/T, \Omega/T)$. 
These data are then recast to the phase transition temperature $T/\mu$ at each $(A/\mu, \Omega/\mu)$.
The derivation of the phase diagram is explained in detail in appendix~\ref{app:DetailPhase}.
The parameter space $(A/\mu,\Omega/\mu)$ can be divided into three regions as in Fig.~\ref{phase0}.
As summarized in section~\ref{intro}, the scalar condensate has 
qualitatively different temperature dependence in each region, and as a consequence the phase transition pattern is also different.
We denote by $T_c^\textrm{1st}$ and $T_c^\textrm{2nd}$ the temperatures for first and second order transitions.
Fig.~\ref{phaseTc} is a contour plot of $T_c^\textrm{1st}$.
There is also a second order superconducting transition in regions A and C at $T_c^\textrm{2nd}$, which has already been computed in Fig.~\ref{fig.Tb}: 
$T_c^\textrm{2nd}=T_\textrm{branch}$ in regions A and C.
We do not draw $T_c^\textrm{2nd}$ in this figure for visibility.
On the border of region B and the others, the temperature of the first order phase transition is smoothly connected to that of the second order one: $T_c^\textrm{1st} = T_c^\textrm{2nd}$ on the border of region B.
The contour plot shows that, at fixed $A$, the phase transition temperature decreases as $\Omega/\mu$ is increased.
Thus, the superconducting phase transition temperature is not increased by the effect of the rotating electric field.
A 3D plot of $T_c^\textrm{1st}$ is given in Fig.~\ref{Tc3D}.
On the upper-right boundary of the region C, the transition temperature becomes zero.
This indicates that the Floquet holographic superconductor admits a first order quantum phase transition.\footnote{For $\langle O_2 \rangle$, the probe limit model can reach $T \to 0$ without a divergence of $\langle O_2 \rangle$~\cite{Hartnoll:2008kx}.}

In the absence of the spatial gauge field ($A=0$), 
the superconducting transition occurs at $T/\mu=0.0587$ (Fig.~\ref{fig.Tb}).
The temperature of the first order transition in Fig.~\ref{phaseTc} is always lower than that value.
This implies that there is no enhancement of superconductivity.
We also checked that, even if we consider the ensemble of a fixed charge density $\rho$,
there is no enhancement of the superconductivity.

\section{Other physical quantities}
\label{sec:other}

\subsection{Current and Joule heating}
\label{sec:joule}

\begin{figure}[t]
\centering
\subfigure[Current $|J|$]{\includegraphics[scale=0.45]{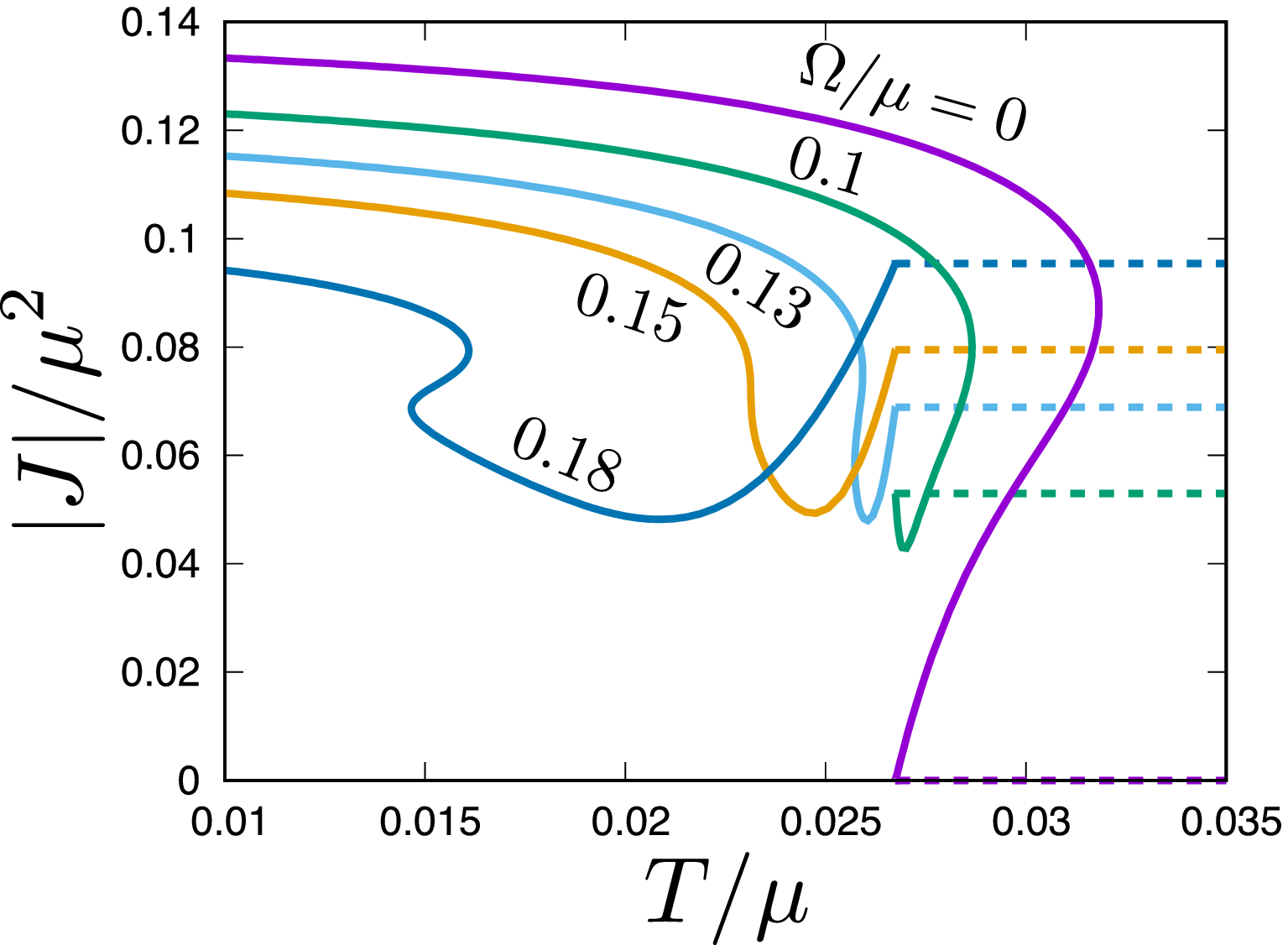}\label{fig:E053_J}}
\subfigure[Joule heating $q$]{\includegraphics[scale=0.45]{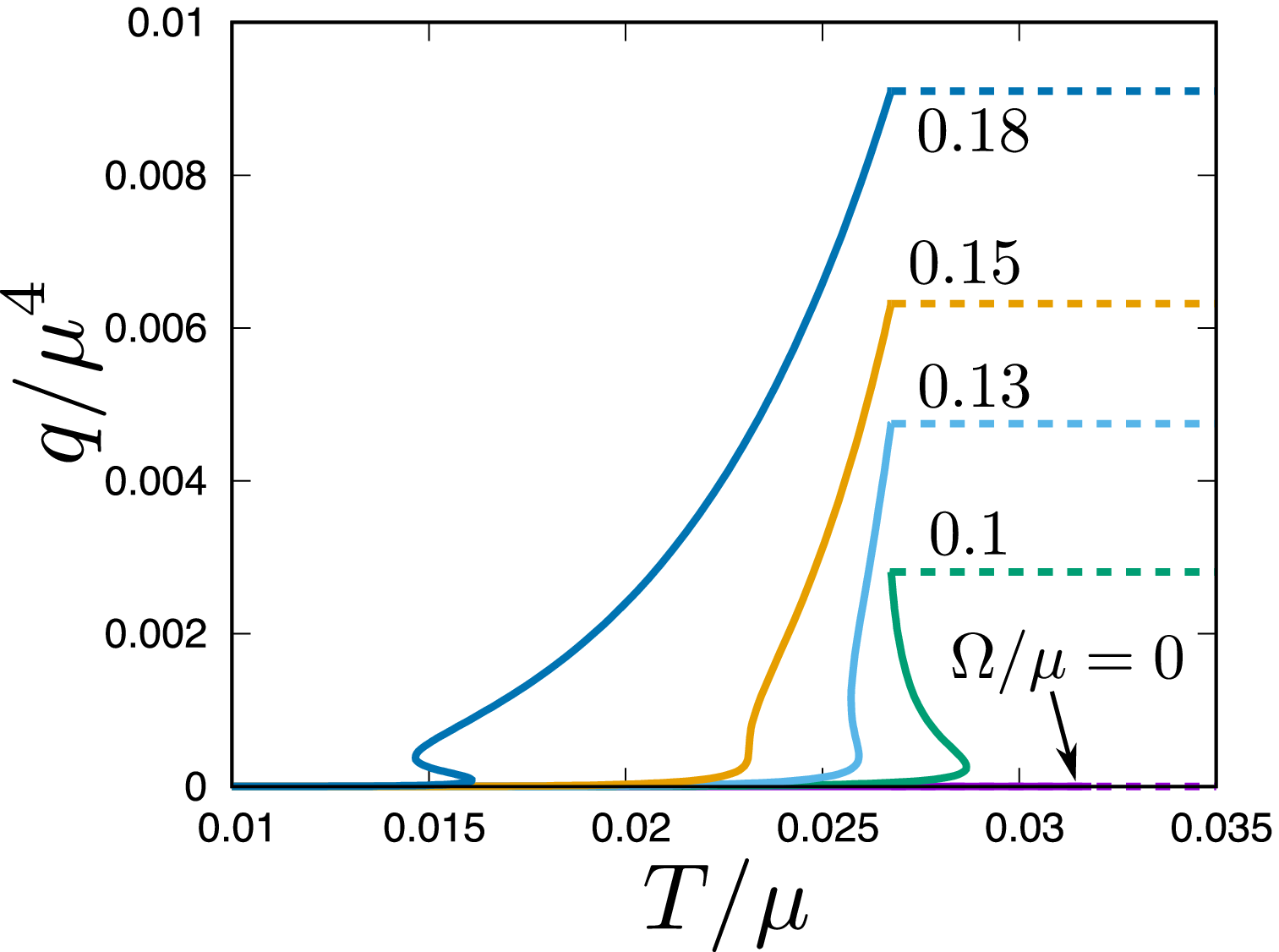}\label{fig:E053_q}}
\caption{Current and Joule heating for $A/\mu=0.53$, $\Omega/\mu=0, \, 0.1, \, 0.13, \, 0.15, \, 0.18$. Dashed lines are normal phase values.}
\label{fig:E053}
\end{figure}

In Fig.~\ref{fig:E053}, we show results of the electric current and Joule heating.
In the plots except $\Omega =0$, the current shows a slight dip around the branching temperature (though it depends on the actual phase transition whether or not the dip can be observed physically, especially for the case of a first order transition), but the current is nonzero in low temperatures.
The Joule heating is suppressed in low temperatures.

\subsection{Normal and superconducting currents}
\label{sec:london}

Since $\vec{\mathcal{A}} \cdot \dot{\vec{\mathcal{A}}}=0$ in our source, the current can be naturally decomposed into the components orthogonal and parallel to the rotating source as
\begin{equation}
\mathcal{J} = c_1 \dot{\mathcal{A}} + c_2 \mathcal{A} \ .
\label{london_JcA}
\end{equation}
With our ansatz, this reduces to
\begin{equation}
J = i c_1 \Omega A + c_2 A \ ,
\end{equation}
and thus the coefficients $c_1, c_2$ are given by
\begin{equation}
c_1 = \frac{1}{\Omega}\mathrm{Im} \left(\frac{J}{A}\right) = -\frac{q}{\Omega^2 |A|^2} \ , \quad
c_2 = \mathrm{Re} \left(\frac{J}{A}\right) \ .
\label{london_c1c2}
\end{equation}
These coefficients can then be compared with those in empirical equations. From \eqref{london_JcA}, we obtain
\begin{equation}
\dot{\mathcal{J}} = - c_1 \dot{\mathcal{E}} - c_2 \mathcal{E} \ .
\end{equation}
This equation contains Ohm's law for normal conductors $\mathcal{J}_n = \sigma_n \mathcal{E}$ and the first London equation $\dot{\mathcal{J}}_s = \rho_s \mathcal{E}$ for superconductors as we identify $\sigma_n = -c_1$ and $\rho_s = -c_2$.
Using \eqref{nomalPhysq}, we obtain $\sigma_n=1$ and $\rho_s=0$ for the normal phase.

\begin{figure}[t]
\centering
\subfigure[Normal component $\sigma_n$]{\includegraphics[scale=0.45]{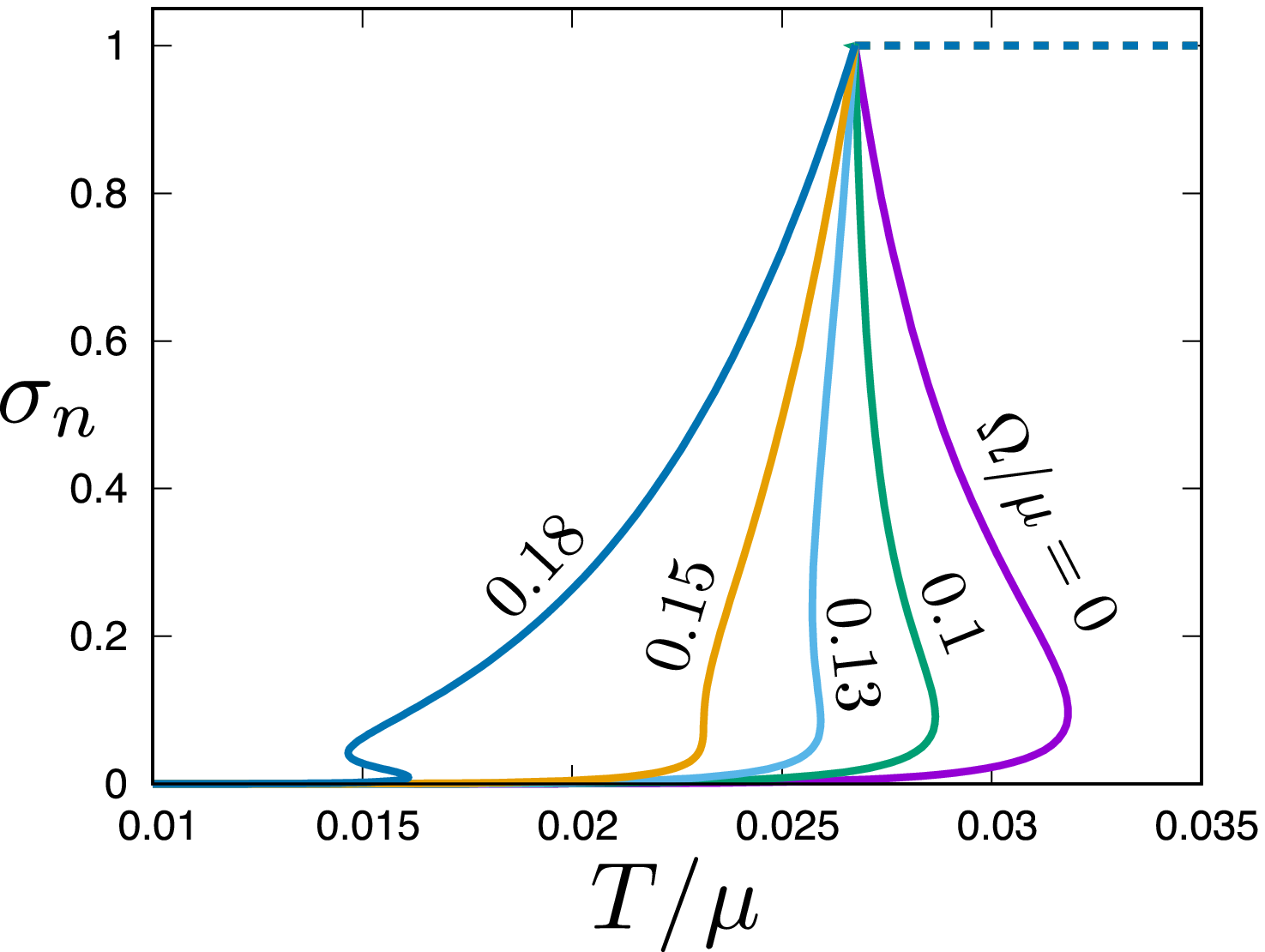}\label{fig:E053_sigma_n}}
\subfigure[Superconducting component $\rho_s$]{\includegraphics[scale=0.45]{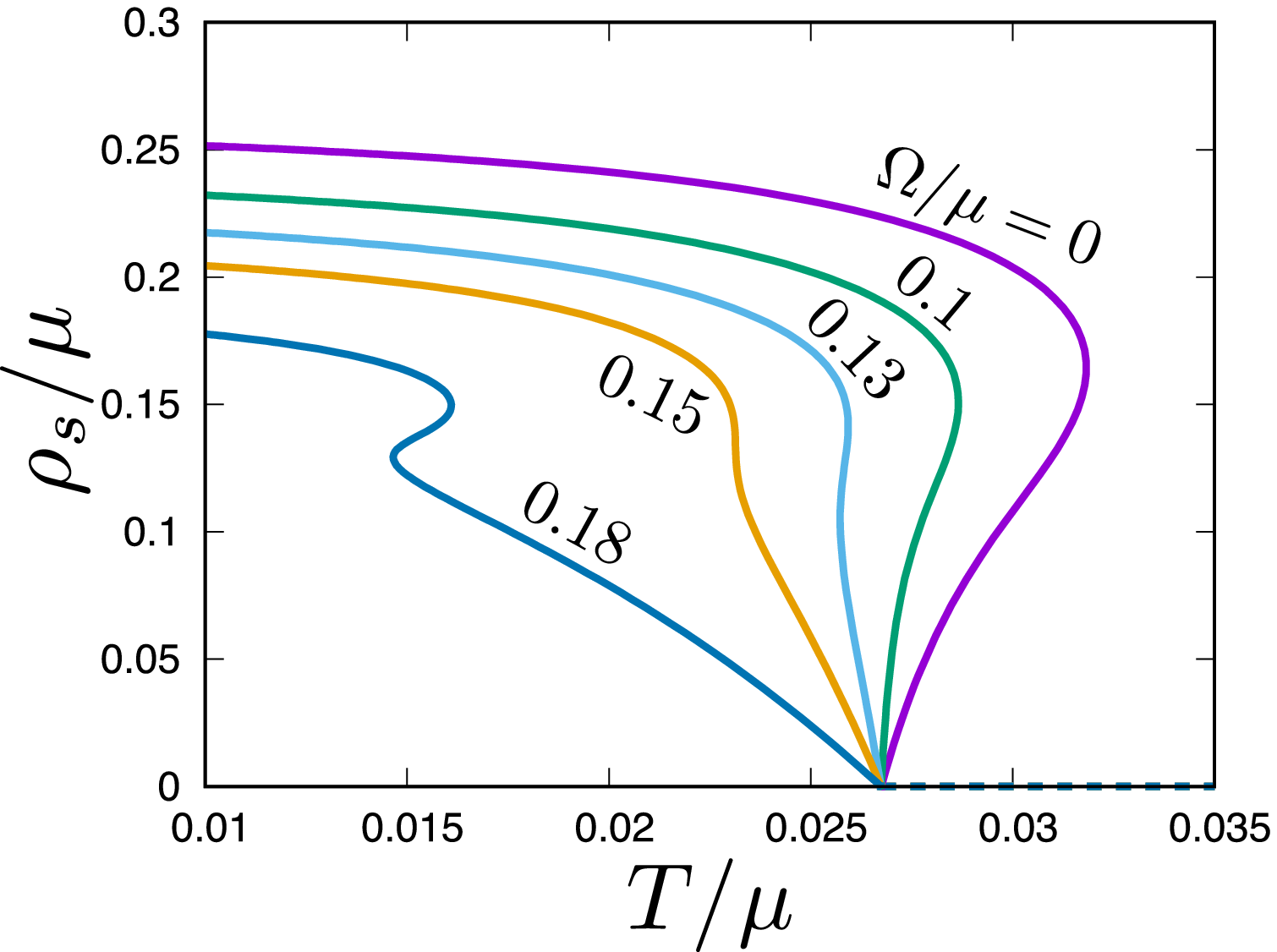}\label{fig:E053_n_s}}
\caption{The conductivity $\sigma_n$ and superfluid density $\rho_s$ for $A/\mu=0.53$ and $\Omega/\mu=0, \, 0.1, \, 0.13, 0.15, \, 0.18$. Horizontal dashed lines in each panel indicate the normal phase values $\sigma_n=1$ and $\rho_s=0$.}
\label{fig:E053_london}
\end{figure}

In Fig.~\ref{fig:E053_london}, we show $\sigma_n$ and $\rho_s$ at $A/\mu=0.53$ and for several $\Omega$.
From $\sigma_n$, we find that the normal component of the current gets suppressed in the superconducting phase.
As $\Omega/\mu$ is increased, the switch from the normal to superconducting currents is slowed down in low temperatures.
For fixed $A/\mu$ so that the transition temperature is the same, a larger $\Omega/\mu$ induces a bigger normal component and a smaller superconducting one.

In Sec.~\ref{NoneqF}, we argued that the breakdown of \eqref{maxcon_integ} appears to stem from the presence of the rotating normal current.
Conversely, when the normal current is quite suppressed, \eqref{maxcon_integ} is satisfied in great accuracy, and therefore the low temperature of the superconducting phase may be well subject to static thermodynamics despite the presence of the time dependent rotating electric field.

\section{Summary}

We studied a holographic superconductor in the presence of a rotating external electric field.
A summary of the resulting phase diagram is given in Fig.~\ref{phase0}.
In particular, we find the parameter region where the scalar condensate develops an ``inverse S-shape,'' and 
there is a first order phase transition inside the condensed phase.
To evaluate the phase transition, we discussed Maxwell construction 
and introduced a reasonable definition of free energy for our rotating electric field setup.
We also considered the decomposition of the induced current to the normal and superconducting components.
The normal component is quite suppressed in the superconducting phase, and 
this suggests that the quite low temperatures can be considered very static because thermodynamic free energy is well posed despite the time periodic driving.

While there appears a new phase structure in the superconducting phase, we did not observe any enhancement of superconductivity by the application of the rotating electric field.
The suppression of superconductivity by an external electric field has been studied in Ref.~\cite{Natsuume:2013lfa}, and the direct effects of the rotating electric field also appear to disfavor the enhancement of superconductivity.
It would remain an interesting question if there are mechanisms to enhance superconductivity in holography.
Since an applied laser perturbs the lattice structure of condensed materials to bring the system to nonequilibrium where superconductivity is enhanced, it might be necessary to introduce spatial inhomogeneity in order to realize similar effects.
In the presence of inhomogeneity, it has been reported that the critical temperature for superconductivity increases compared to homogeneous configurations~\cite{Ganguli:2012up,Horowitz:2013jaa,Arean:2013mta}.
Such increased superconductivity may be further enhanced in nonequilibrium states by driving by an external electromagnetic field.

We used the probe approximation for the study of the Floquet holographic superconductor.
For the reduction to the one-dimensional problem, this approximation is essential:
If we take into account the backreaction, the black hole spacetime cannot be stationary anymore 
because of the constant energy flow from the boundary to the horizon~\cite{Biasi:2017kkn}.
However, even for the backreacted case, 
our probe analysis would give a good approximation 
while the total injected energy is much smaller than the initial black hole mass.

In this paper, we focused only on spatially homogeneous solutions. It will be important to address possible dynamical instabilities.
A next step will be to study the quasinormal modes in the holographic superconductor~\cite{Amado:2009ts}. When $\Omega=0$ and $A \neq 0$, finite momentum instabilities are found in homogeneous superconducting solutions~\cite{Amado:2013aea}. Finite $\Omega$ will be straightforwardly included in the analysis of quasinormal modes. If such instabilities exist, it may nonperturbatively result in the formation of superconducting vortices and turbulence~\cite{Adams:2012pj,Dias:2013bwa,Chesler:2014gya,Ewerz:2014tua,Du:2014lwa,Lan:2016cgl}.

If we set $A=0$, we do not obtain any self-maintained Floquet state with spontaneous current $J\neq0$ analogous to boson stars~\cite{Astefanesei:2003qy,Buchel:2013uba} or Floquet condensation states~\cite{Kinoshita:2017uch}. This is because of the existence of the horizon acting as a dissipator. There will be a chance to construct such Floquet solutions with spontaneous currents in dissipationless holographic backgrounds including confining backgrounds and AdS solitons. 

The linear perturbation of the Floquet holographic superconductor is an important direction for future work. We can study the dynamical stability of states in the multivalued region of the order parameter by the perturbation analysis. 
The DC and AC conductivities can also be computed in a similar way as in Ref.\cite{Hashimoto:2016ize}. Since the rotating electric field explicitly violates the spacial parity, the Hall current would be induced in the current system. 

The circularly polarized laser is used in experiment 
to realize the Floquet topological insulator~\cite{Wang13}.
On the other hand, the superconductivity 
in the calcium-intercalated bilayer graphene is experimentally realized~\cite{Ichinokura}.
It would be nice if we can verify our holographic results 
through the development of such works on superconductivities and Floquet states.

\acknowledgments
The authors would like to thank Philippe Sabella-Garnier, Umut G\"{u}rsoy, Hisao Hayakawa, Matti J\"{a}rvinen, Aurelio Romero-Berm\'{u}dez, and Jan Zaanen for useful conversation.
We would particularly like to thank Shin Nakamura for useful comments on the steady state thermodynamics in AdS/CFT.
We would also like to thank organizers and participants of ``International Molecule Program on Floquet Theory : Fundamentals and Applications'' 
for the opportunity to present this work and useful comments.
We would particularly like to thank Takashi Oka for kindly informing us about the theory and experiment of the enhancement of superconductivity. 
The work of T.~I.~was supported by the Netherlands Organisation for Scientific Research (NWO) under the VIDI grant 680-47-518 and the Delta-Institute for Theoretical Physics ($\Delta$-ITP), which is funded by the Dutch Ministry of Education, Culture and Science (OCW).
The work of K. M. was supported by JSPS KAKENHI Grant No. 15K17658 and 
in part by JSPS KAKENHI Grant No. JP17H06462.
\appendix

\section{Numerical details for superconducting solutions}
\label{Numerics}

In this appendix, we explain details of numerically solving the bulk equations 
in the superconducting phase $\Psi\neq 0$.
For numerical calculations, 
we introduce $\chi\equiv \Psi/z$, 
which behaves as $\chi=\psi_1 + \psi_2 z+\cdots$ near the AdS boundary.
We also use the tortoise coordinate $r_\ast$ instead of $z$ so as to keep resolution in the near horizon region.
The equations of motion for $\chi(r_\ast)$, $b(r_\ast)$ and $A_t(r_\ast)$ are simply written as
\begin{equation}
\begin{split}
&\chi_{,r_\ast r_\ast}=
\left[
f|b|^2-A_t^2
+\frac{zf}{z_h^3}
\right]\chi
\ ,\\
&b_{,r_\ast r_\ast}
=\left(-\Omega^2 
+2f \chi^2\right)b\ ,\\
&A_t{}_{,r_\ast r_\ast}= -f' A_t{}_{,r_\ast}
+2f\chi^2 A_t \ .
\end{split}
\label{EOMrast}
\end{equation}
These equations are regular at $z=0$. 
Therefore, we need not introduce any cutoff near the AdS boundary for numerical integration.
In our numerical calculations, we set $z_h=1$ ($T=3/4\pi$) using the scaling symmetry~(\ref{scalingsym}).
We can choose $b_H$ real since a shift of its phase, $b_H\to b_H e^{i\theta}$, 
just changes the time independent phases of the electric field and current as $A\to A e^{i\theta}$ and $J\to Je^{i\theta}$.
Once numerical solutions are obtained, we can use this phase rotation to make $A$ real as we do in the figures. 
Then, the bulk solutions are parametrized by 4 real parameters $(\Psi_H,b_H,A_H,\Omega)$.
By the 4th order Runge-Kutta method, 
we numerically integrate Eq.~(\ref{EOMrast}) 
from the horizon (for which we use a numerical cutoff $r_\ast\simeq -3.0$) to the AdS boundary ($r_\ast = 0$).
One of the purposes of the numerical calculations is to figure out 
$(T/\mu)$-dependence of $\sqrt{\langle O_2 \rangle}/\mu$ for fixed $A/\mu$ and $\Omega/\mu$ as in Fig.~\ref{O2}.
When we construct the data for the figure, we fix $\Psi_H$ and 
tune 3 parameters $(b_H,A_H,\Omega)$ by the shooting method 
so that $\psi_1=0$ is satisfied and 
($A/\mu$, $\Omega/\mu$) become desired values.
Once we find a preferred solution by the shooting, 
we read out the physical quantities ($\langle O_2\rangle$, $\mu$, $\rho$, $A$, $J$, $q$) from the asymptotic form of the solution near the AdS boundary.
From these quantities, we compute scaling invariant combinations such as $\sqrt{\langle O_2 \rangle}/\mu$ and $T/\mu$.
We change $\Psi_H$ as $\Psi_H=(c_1+c_2n)^4$ for $n=0,1,2,\cdots$. 
(Since the physical quantities change rapidly for small $\Psi_H$, we took small steps for small $\Psi_H$.)
Typically, we choose $c_1\simeq 0.1$ and $c_2\simeq 0.01$.
For $n=0$, since $\Psi_H$ is small, 
we can use the normal phase solution as the initial guess of $(b_H,A_H,\Omega)$.
For the initial guess of the $n$th $(b_H,A_H,\Omega)$, we use the $(n-1)$th values. 
Repeating this procedure, 
we can find the $(T/\mu)$-dependence of $\langle O_2 \rangle/\mu^2$. 

\section{Detailed derivation of the phase structure}
\label{app:DetailPhase}

In this appendix, we explain how to derive the phase structure of the Floquet holographic superconductor 
(Figs.~\ref{phase0} and~\ref{phaseTc}).
To determine the transition temperature, we use the Maxwell construction discussed in section~\ref{Maxwell}.
For the Maxwell construction, we need to know the $\mu$-dependence of $\rho$ 
for fixed ($A$, $\Omega$, $T$). Using the scaling symmetry~(\ref{scalingsym}), 
we will non-dimensionalize the physical quantities by the temperature $T$ in this appendix.
For numerical calculations, 
we basically use the same method as in appendix~\ref{Numerics}, 
but we use ($A/T$, $\Omega/T$) as the shooting target parameters instead of ($A/\mu$, $\Omega/\mu$).
For fixed ($A/T$, $\Omega/T$), 
the horizon value of the scalar field is varied as $\Psi_H=(c_1+c_2n)^4$ for $n=0,1,2,\cdots$ ($c_1\simeq 0.1$, $c_2\simeq 0.01$).
For each $n$, we obtain 
the charge density $\rho=\rho_n$ and chemical potential $\mu_n$. From the discrete data, we compute the ``free energy'' 
at $\mu=\mu_n$ as
\begin{equation}
 F(\mu_n)=-\sum_{m=1}^n \frac{\rho_m+\rho_{m-1}}{2}(\mu_m-\mu_{m-1})\ .
\end{equation}
In the normal phase, the free energy is analytically given by 
$F_\textrm{normal}(\mu)=-2\pi T (\mu^2-\mu_0^2)/3$. 
We introduce the difference of the free energy from the normal phase
\begin{equation}
 \delta F(\mu)=F(\mu)-F_\textrm{normal}(\mu)\ .
\end{equation}

\begin{figure}
  \centering
  \subfigure[$A/T=20$, $\Omega/T=2$]
 {\includegraphics[scale=0.45]{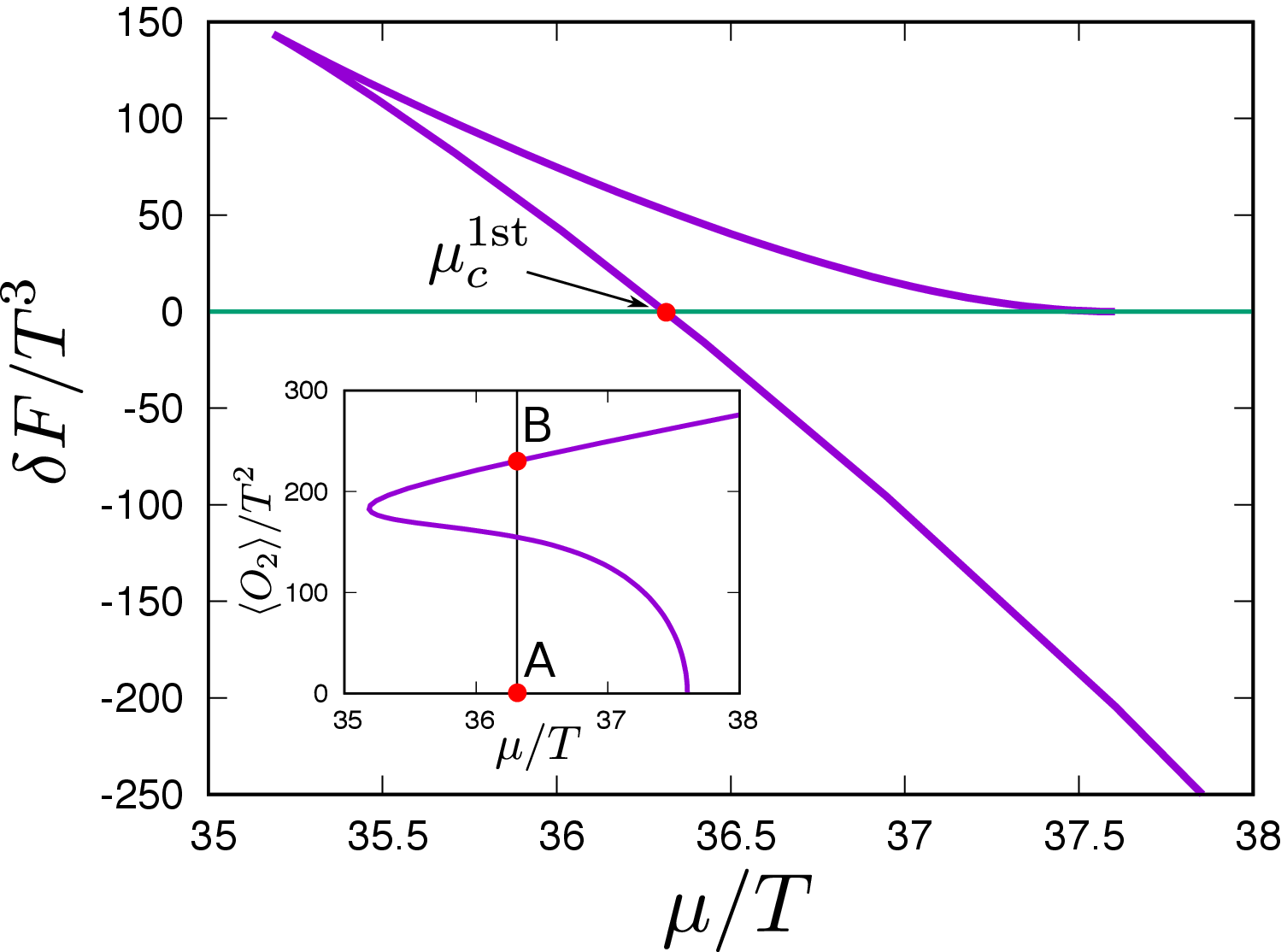}\label{F1st}
  }
\subfigure[$A/T=30$, $\Omega/T=6$]
 {\includegraphics[scale=0.45]{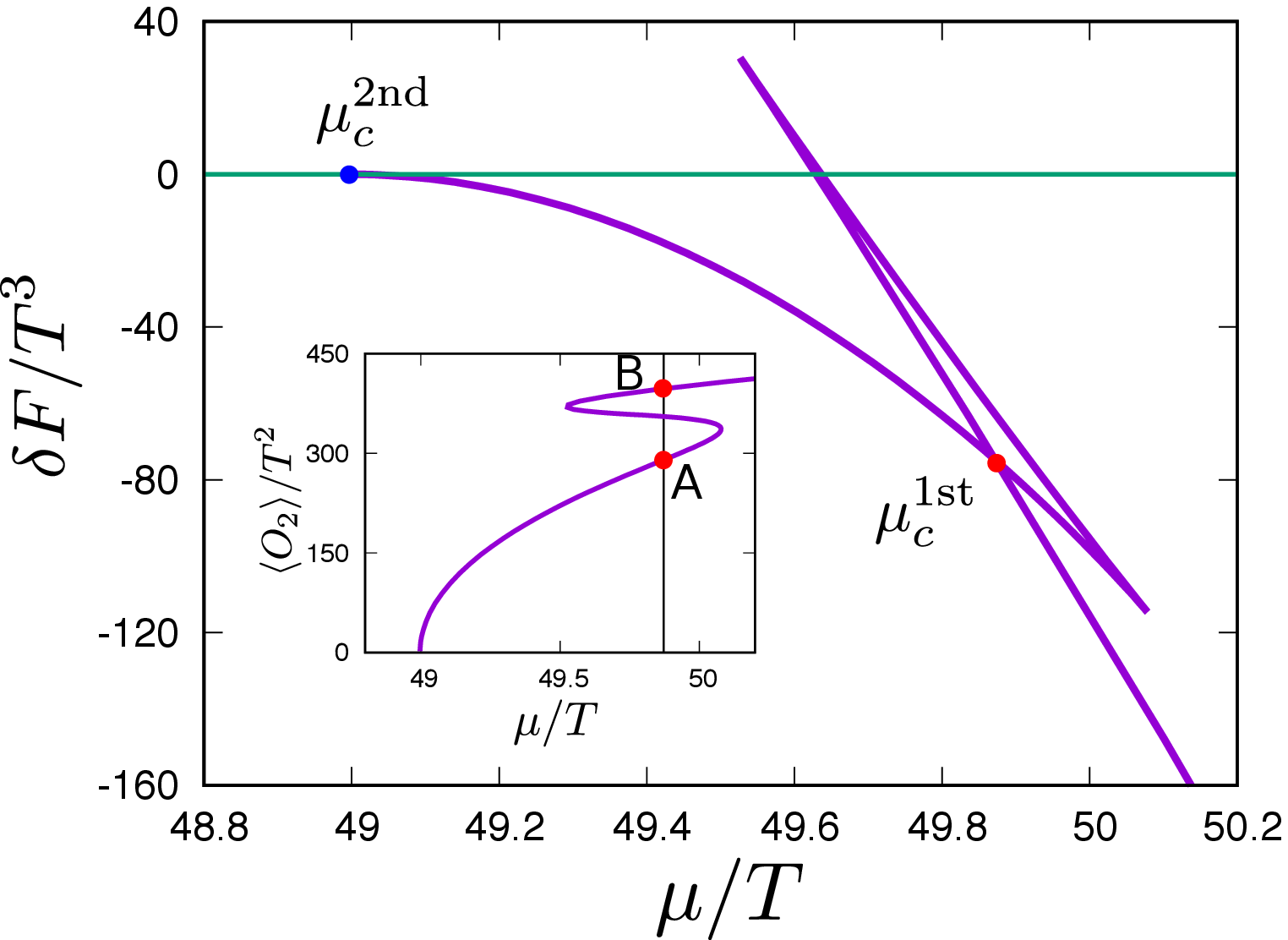}\label{F2nd1st}
  }
 \caption{
Typical profiles of the free energy measured from the normal phase.
\label{Ftypical}
}
\end{figure}

Fig.~\ref{Ftypical} shows examples of $\delta F(\mu)$.
We set $(A/T, \Omega/T)=(20,2)$ for panel (a) and 
$(A/T, \Omega/T)=(30,6)$ for panel~(b).
We denote critical values of chemical potential for first and second order phase transitions by 
$\mu_c^\textrm{1st}$ and $\mu_c^\textrm{2nd}$, respectively.
In panel~(a), 
 $\delta F$ intersects $\delta F=0$ at $\mu_c^\textrm{1st}/T=36.3$.
This indicates a first order phase transition.
In panel~(b), the superconducting phase branches from the normal phase at $\mu_c^\textrm{2nd}/T=49.0$. 
This is a second order superconducting phase transition.
We also find a self-intersection of the curve at $\mu_c^\textrm{1st}/T=49.9$.
This implies that there is 
a first order transition between superconducting phases.
In insets, we show superconducting order parameters $\langle O_2\rangle$  as functions of $\mu$.
In each panel, the first order transition point $\mu=\mu_c^\textrm{1st}$ is shown in vertical lines, and 
there is a phase transition between points A and B.

\begin{figure}
  \centering
  \subfigure[$\mu_c^\textrm{1st}$]
 {\includegraphics[scale=0.55]{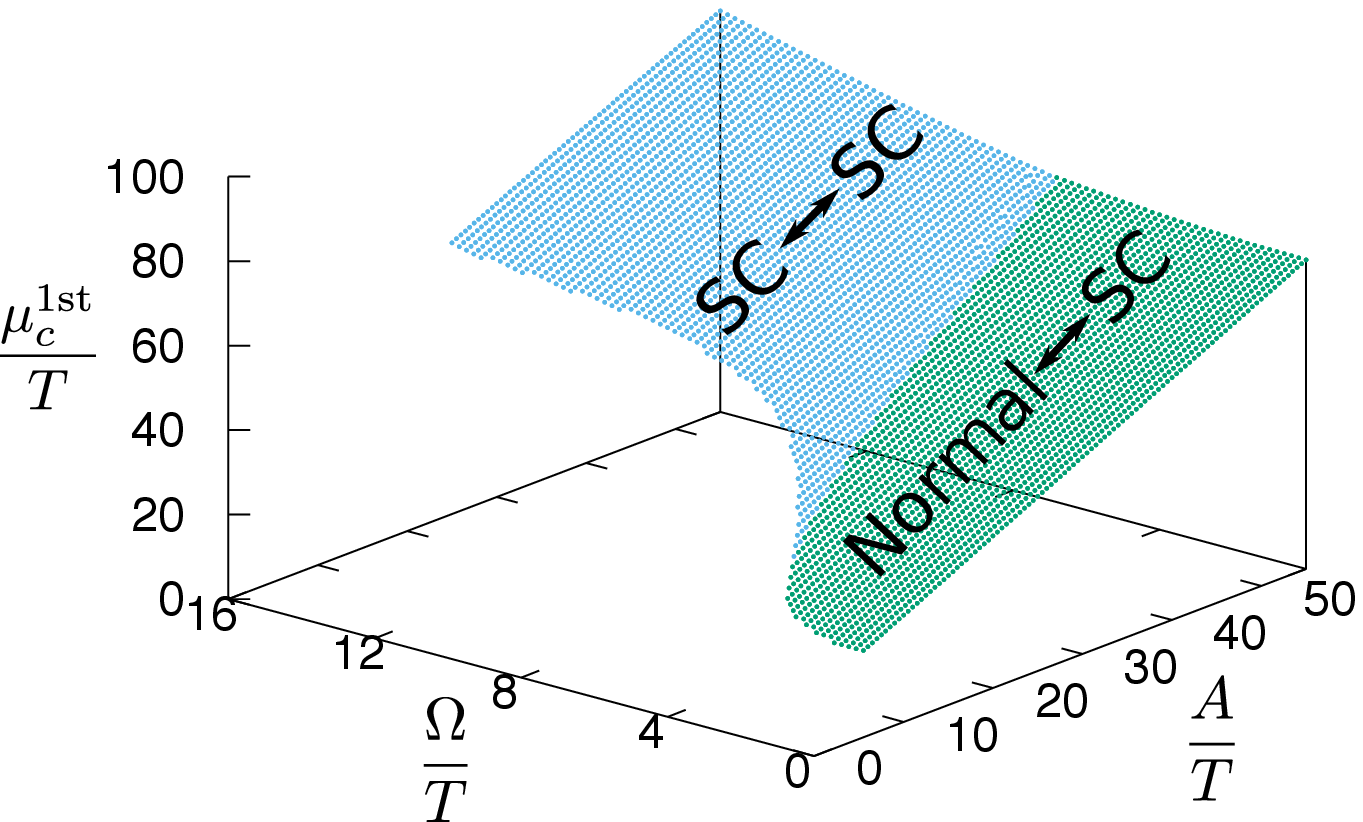}\label{muc3D}
  }
\subfigure[$T_c^\textrm{1st}$]
 {\includegraphics[scale=0.55]{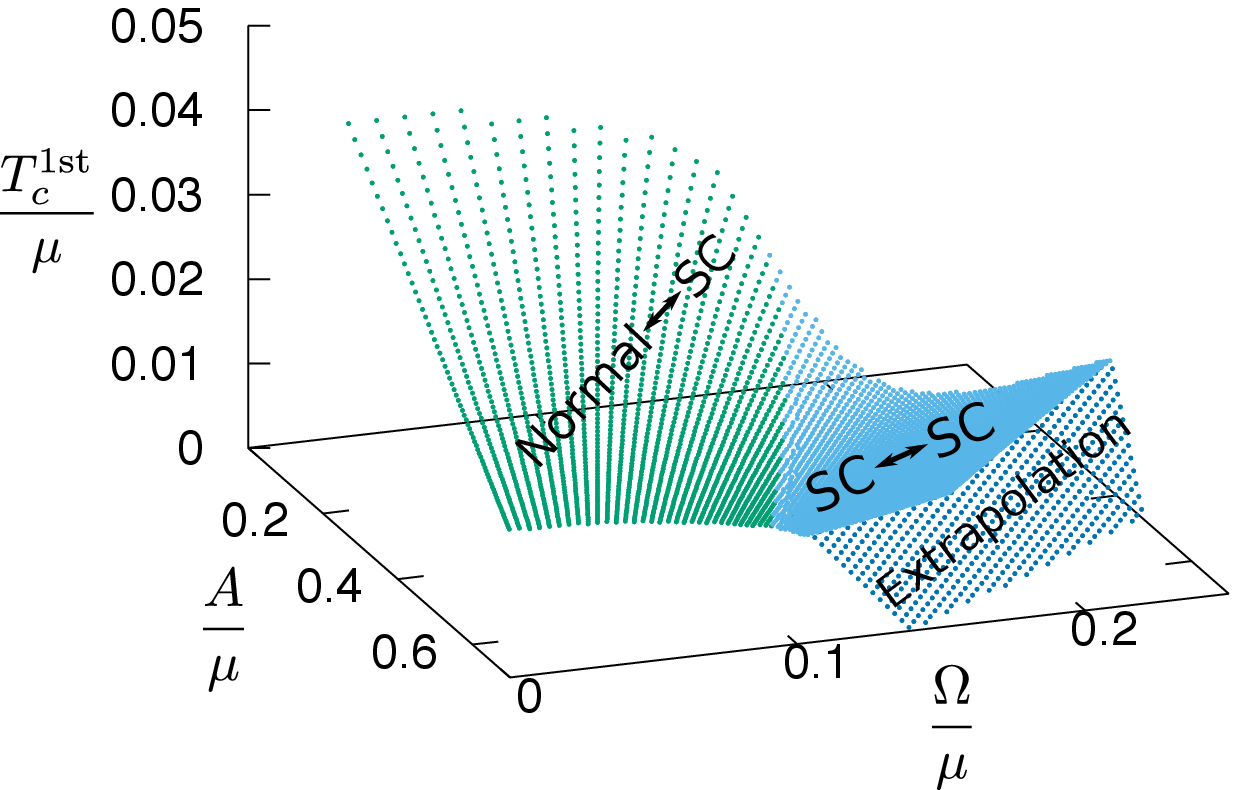}\label{Tc3D}
  }
 \caption{
Critical chemical potential and temperature for the first order phase transition.
\label{critmuc}
}
\end{figure}

We repeat the calculation described above for 
$(A/T,\Omega/T)=(0.5i,0.2j)$ where $i=1,2,\cdots,110$ and $j=0,1,\cdots,80$.
For each parameter, 
we obtain the critical value of the first order transition $\mu_c^\textrm{1st}$.
Fig.~\ref{muc3D} shows a 3D plot of $\mu_c^\textrm{1st}$ as a function of $A$ and $\Omega$.
We do not show $\mu_c^\textrm{2nd}$ because it is obtained by perturbative analysis in section~\ref{perturbation}.
Green points correspond to transitions between normal and superconducting (SC) phases (as in Fig.~\ref{F1st}).
Light blue points are transitions within superconducting phases (as in Fig.~\ref{F2nd1st}).
We can find that the light blue sheet is almost planar in the region of large $A$ and $\Omega$.
The light blue sheet is well approximated by
\begin{equation}
 \frac{\mu_c^\textrm{1st}}{T}=1.01\times \frac{A}{T}+2.09\times \frac{\Omega}{T}+5.76\ .
\label{Extrapolate1}
\end{equation}
We can expect that the critical chemical potential $\mu_c^\textrm{1st}$ is given by the above expression
even outside of our numerical domain.
From the numerical data of ($A/T$, $\Omega/T$, $\mu_c^\textrm{1st}/T$),
we can generate data of the transition temperature in the grand canonical ensemble as
$(A/\mu$, $\Omega/\mu$, $T_c^\textrm{1st}/\mu)
=(A/T \times T/\mu_c^\textrm{1st}$, $\Omega/T \times T/\mu_c^\textrm{1st}$, $T/\mu_c^\textrm{1st})$.
The result is shown in Fig.~\ref{Tc3D}.
The region labeled ``Extrapolation'' is outside of our numerical domain, 
but we generate the data from Eq.~(\ref{Extrapolate1}) as
\begin{equation}
 \frac{T_c^\textrm{1st}}{\mu}=0.173-0.175\times \frac{A}{\mu}-0.363\times \frac{\Omega}{\mu}\ .
\label{Extrapolate2}
\end{equation}
The contour plot of Fig.~\ref{Tc3D} is nothing but Fig.~\ref{phaseTc}.
Looking at this figure from above (i.e. projecting data onto a horizontal plane), 
we obtain Fig.~\ref{phase0}. 
The curves of the boundaries of regions A, B and C 
are constructed by interpolating/extrapolating discrete numerical data.

\section{Integrability condition involving a scalar source}
\label{app:scalar}

In this appendix, we evaluate the integrability condition when $\delta \psi_1$ is also taken into account.
Restoring the $\delta \psi_1$ variation and carrying out holographic renormalization, the renormalized on-shell action \eqref{s_ren_variation} is modified to
\begin{equation}
s_\infty = \int dt d^2x \left[\rho \delta \mu + \vec{\mathcal{J}}(t)\cdot \delta \vec{\mathcal{A}}(t) + \psi_2^\ast \delta \psi_1 + \psi_2 \delta \psi_1^\ast \right]\ .
\end{equation}
Without loss of generality, we can set $\psi_{1,2}$ real.
Then, the integrability of the sources and responses also requires
\begin{equation}
\frac{\partial \rho}{\partial \, \Xi_1} = \frac{\partial \langle O_2 \rangle}{\partial \mu} \ ,\quad
\frac{\partial \langle O_2 \rangle}{\partial \vec{A}}=\frac{\partial \vec{J}}{\partial \, \Xi_1} \ ,
\label{integcond_phi}
\end{equation}
as well as \eqref{maxcon_integ}, where $\Xi_1 = \sqrt{2} \psi_1$ is the scalar source.

\begin{figure}[t]
\centering
\includegraphics[scale=0.45]{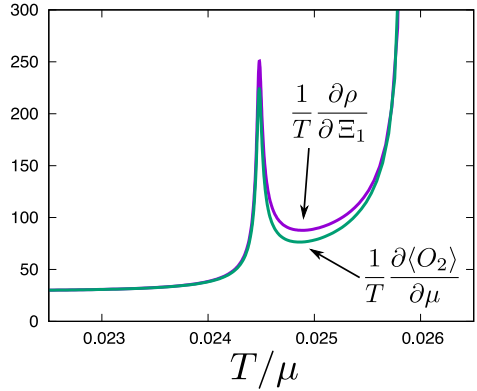}
\caption{
Violation of the integrability conditions, evaluated when $A/T=20.94$, $\Omega/T=6.283$ and $\Xi_1=0$, which are the same parameters as in Fig.~\ref{fig:integ_vio}. The $\Xi_1$-derivative is calculated around the trivial scalar source background.
The divergence to the right is due to the superconducting phase transition and therefore is inessential.
}
\label{fig:integ_vioP}
\end{figure}

When $\Omega=0$, we check that the integrability is satisfied as well.
When $\Omega \neq 0$, again, the integrability conditions are violated as shown in Fig.~\ref{fig:integ_vioP}.\footnote{We obtain qualitatively the same violation also in the case of a nonzero scalar source, $\Xi_1 \neq 0$.}
The violation of the first relation in \eqref{integcond_phi} implies that a function $F$ satisfying even $dF=-\rho d\mu-\langle O_2 \rangle d\Xi_1$ does not exist, which involves the static sources $\mu$ and $\Xi_1$ rather than the dynamical $\vec{\mathcal{A}}$.
This observation appears to indicate a deep trouble in constructing a consistent thermodynamic potential in nonequilibrium systems.
In our setup in the main text, we have only $\mu$ as a static source, and we use it to define the free energy~(\ref{Fintrhodmu}).

We are hence aware that \eqref{Fintrhodmu} is not the unique construction to evaluate the quantitative value of, e.g., the phase transition temperature.
The same concern would apply to general studies of the phase structures of steady solutions in the presence of a dynamical source.
We intend that our phase transition search aims to find qualitatively definite features.

\end{document}